\documentstyle[12pt]{article}
\begin{document}
\font\helv=phvr at 12pt
\def\ghost#1#2#3#4#5#6 {\raise #1 pt \hbox{$
\stackrel{(#2,#3)}{{#4}_{{#5}_{#6}}}$}}
\def\ghostsp#1#2#3#4#5#6#7 {\raise #1 pt \hbox{$
\stackrel{(#2,#3)}{{#4}^{#5}_{{#6}_{#7}}}$}}
\def\cghost#1#2#3#4#5#6 {\raise #1 pt \hbox{$
\stackrel{(#2,#3)}{{#4}^{{#5}_{#6}}}$}}

\begin{titlepage}
\hspace{9cm} ULB--PMIF--92/06

\vspace{3cm}
\begin{centering}

{\Large Hamiltonian BRST-anti-BRST Theory}
\vspace{1cm}

{\large Philippe Gr\'egoire$^\dagger$ and Marc Henneaux$^*$}\\

Facult\'e des Sciences, Universit\'e Libre de Bruxelles,\\
Campus Plaine C.P. 231, B-1050 Bruxelles, Belgium\\
\vspace{2cm}
{\large Abstract}
\vspace{.5cm}

\end{centering}

The hamiltonian BRST-anti-BRST theory is developed in the general case
of
arbitrary reducible first class systems. This is done by extending the
methods of
homological perturbation theory, originally based on the use of a
single
resolution, to the case of a biresolution. The BRST and the anti-BRST
generators
are shown to exist. The respective links with the ordinary BRST
formulation
and with the $ sp(2) $-covariant formalism are also established.

\vspace{2cm}
\vspace{1cm}

\noindent
{\footnotesize ($^\dagger$)Chercheur IRSIA. \\
($^*$)Ma{\^\i}tre de recherches au Fonds National de la Recherche
Scientifique. Also
at Centro de Estudios Cientificos de Santiago, Casilla 16443. Santiago
9, Chile.}

\thispagestyle{empty}
\vfill
\end{titlepage}
\pagebreak

\section{Introduction}
It has been realized recently that the proper algebraic setting for the
BRST
theory is that of homological perturbation theory
\cite{Stasheff,Fish:Henneaux:Stasheff:Teitelboim}.
Homological perturbation theory permits one not only to prove the
existence of the
BRST transformation, both in the lagrangian and the hamiltonian cases,
but also
establishes that the BRST cohomology at ghost number zero is given by
the physical
observables (the gauge invariant functions). These key properties,
valid for irreducible or reducible gauge theories with closed or
``open"
algebras are what make
the BRST formalism of physical interest
\cite{Fish:Henneaux:Stasheff:Teitelboim,Fisch:Henneaux,DuboisViolette,H
enTeit,Henneaux:Teitelboim}.

The purpose of this paper is to extend the analysis of
\cite{Fish:Henneaux:Stasheff:Teitelboim}
to cover the anti-BRST transformation.
The anti-BRST symmetry was formulated in the context of Yang-Mills
theory
immediately after the BRST symmetry was discovered
\cite{Curci:Ferrari,Ojima}. Altough it does not play a
role as fundamental as the BRST symmetry itself, it is a useful tool in
the
geometrical (superfield) description of the BRST transformation, in the
investigation of perturbative renormalizibility of Yang-Mills models,
as well as
in the understanding of the so-called non minimal sector
\cite{Hwang,Bonora:Tonin,Bonora:Pasti:Tonin,BauThier,Ore:van
Nieuwenhuizen,Alvarez-Gaume:Baulieu,Hwang2}.
For all these reasons, it is of interest to develop the BRST-anti-BRST
formalism
in the general case of an arbitrary gauge system.

We show in this article that the methods of homological perturbation
theory can
be adapted to cover the anti-BRST transformation. This is done by
duplicating
each differential appearing in the BRST construction. In particular,
the crucial
Koszul-Tate complex \cite{Fish:Henneaux:Stasheff:Teitelboim,Mullan} is
replaced
by the Koszul-Tate bicomplex. The usual existence and uniqueness
theorems for
the BRST generator can then be extended without difficulty to the BRST-
anti-BRST
algebra by following the same lines as in the BRST case. Our results
were
announced in \cite{Gregoire:Henneaux}.

Although we consider here only the hamiltonian method, our approach can
also be
applied to the antifield formalism. However, the explicit form of the
biresolution is then different, so that we shall reserve the discussion
of the
antifield anti-BRST theory for a separate publication
\cite{Gregoire:Henneaux2}
\footnote{The lagrangian BRST-anti-BRST formalism has been considered
recently
from different viewpoints in
\cite{Spiridonov,BatLav1,BatLav2,Gomis:Roca,Gates,Hull-Spence,Hull}.}.

Our paper is organized as follows. In the next section, we briefly
review the
salient facts of homological perturbation theory in the context of the
BRST
symmetry. We then introduce the concept of biresolution and develop its
properties (section \ref{Sec3}). Section \ref{Sec4} is devoted to the
proof of
the main result of this paper, namely the existence of a Koszul-Tate
biresolution associated with any constraint surface $ \Sigma $ embedded
in phase
space. In section \ref{Sec5}, we prove the existence and uniqueness of
the
BRST and the anti-BRST generators. We then establish some results about
the
BRST and the anti-BRST cohomologies (section \ref{Sec6}). Section
\ref{Sec7}
is devoted to the comparison between the BRST-anti-BRST formalism and
the
standard BRST theory; as a byproduct of this comparison, the
equivalence between
the two formulations is proven. In section \ref{Sec8}, we make the
comparisons with the hamiltonian $ Sp(2) $-formalism of references
\cite{BatLav3,BatLav4}.

\section{Homological perturbation theory in brief}
\label{Sec2}
\subsection{Geometrical ingredients of a gauge theory}
\label{Subsec21}
In either the lagrangian or hamiltonian versions, the description of a
gauge
theory involves the following geometrical data :
\begin{enumerate}
\item A smooth manifold $ \Gamma $ with coordinates $ z^{I} $. These
are either
the canonical coordinates of the hamiltonian formalism, or the
``coordinates" of
the histories of the fields in the lagrangian case.
\item A submanifold $ \Sigma \subset \Gamma $ defined by implicit
equations
\begin{equation}
G_{A_{0}}(z^{I}) \approx 0, \qquad A_{0} = 1, \ldots , M_{0}.
\end{equation}
These are the hamiltonian constraints or the Euler-Lagrange equations.
\item A distribution $ \{{\bf X}_{\alpha_{0}}; \alpha_{0} = 1, \ldots,
m_{0} \} $
tangent to $ \Sigma $ and in involution on it :
\begin{equation}
\label{tang}
{\bf X}_{\alpha_{0}}[G_{A_{0}}] \approx 0 ,
\end{equation}
\begin{equation}
\label{invol}
[{\bf X}_{\alpha_{0}},{\bf X}_{\beta_{0}}] \approx C_{\alpha_{0}
\beta_{0}}^{\gamma_{0}} {\bf X}_{\gamma_{0}}.
\end{equation}
The vector fields $ {\bf X}_{\alpha_{0}} $ generate the infinitesimal
gauge
transformations. These map $ \Sigma $ on itself (equation (\ref{tang}))
and are
integrable on $ \Sigma $ (equation (\ref{invol})). The corresponding
integral
submanifolds are the gauge orbits.
\end{enumerate}

The observables of a gauge theory are the functions on $ \Sigma $ that
are
constant along the gauge orbits (gauge invariance). Thus, if we denote
by
$ \Sigma/{\cal G} $ the ``reduced" space obtained by taking the
quotient of
$ \Sigma $ by the gauge orbits, the algebra of observables is just
$ C^{\infty}(\Sigma/{\cal G}) $. In principle, all the physical
information
about the gauge system is contained in $ C^{\infty}(\Sigma/{\cal G}) $.

\subsection{BRST differential}
\label{Subsec22}
In practice, one cannot construct explicitly the algebra
$ C^{\infty}(\Sigma/{\cal G}) $ of physical observables, either because
one
cannot solve the equations defining $ \Sigma $, or because the
integration
of the gauge orbits is untractable. The BRST construction reformulates
the
concept of observables in an algebra that is more convenient, as the
elements
of the zeroth cohomology group of the BRST differential $ s $,
\begin{equation}
\label{nils}
s^{2} = 0.
\end{equation}

Corresponding to the two ingredients contained in the definition of the
observables, namely the restriction to $ \Sigma $ and the condition of
gauge
invariance, there are actually two differentials hidden in $ s $. The
first one
is known as the Koszul-Tate differential $ \delta $ and implements the
restriction to $ \Sigma $. More precisely, it yields a resolution of
the
algebra $ C^{\infty}(\Sigma) $. The second one is (a model for) the
longitudinal
exterior derivative along the gauge orbits and is denoted by $ D $. It
imposes
the condition of gauge invariance. One has
\cite{Stasheff,Fish:Henneaux:Stasheff:Teitelboim,DuboisViolette,Henneau
x:Teitelboim,Forger:Kellendonk}
\begin{equation}
\label{s=delta+}
s = \delta + D + ``{\rm more}"
\end{equation}
and
\begin{equation}
\label{scohomology}
H^{0}(s) \simeq C^{\infty}(\Sigma/{\cal G}).
\end{equation}
The existence of the additional terms in (\ref{s=delta+}) necessary for
the
nilpotency (\ref{nils}) of $ s $ is a basic result of homological
perturbation
theory. It follows from the resolution property of the Koszul-Tate
differential.
We shall not reproduce the proof here but shall rather refer to the
monograph \cite{Henneaux:Teitelboim}.

The equation (\ref{scohomology}) provides the basic link between gauge
invariance and BRST invariance. It explains why the BRST symmetry is
physically relevant.

\section{Biresolutions}
\label{Sec3}
\subsection{Motivations}
\label{Subsec31}
In the BRST-anti-BRST theory, the differential $ s $ is replaced by two
differentials $ s_{1} $ (BRST differential) and $ s_{2} $ (anti-BRST
differential) that anticommute,
\begin{equation}
\label{BRST-anti-BRSTdef}
s_{1}^{2} = s_{1}s_{2} + s_{2}s_{1} = s_{2}^{2} = 0.
\end{equation}
The relations (\ref{BRST-anti-BRSTdef}) define the BRST-anti-BRST
algebra.
Furthermore, both $ s_{1} $ and $ s_{2} $ are such that
\begin{equation}
\label{BRST-anti-BRSTcohomology}
H^{0}(s_{1}) \simeq H^{0}(s_{2}) \simeq C^{\infty}(\Sigma/{\cal G})
\end{equation}
in a degree that will be made more precise below. This suggests that
one should
introduce two resolutions $ \delta_{1} $ and $ \delta_{2} $ of
$ C^{\infty}(\Sigma) $ that anticommute, instead of the single Koszul-
Tate
resolution $ \delta $ of the BRST theory. Thus, we are led to the
concept of
biresolution.

\subsection{Definitions}
\label{Subsec32}
\newtheorem{biresolution}{Definition}[section]
Let $ {\cal A}_{0} $ be an algebra and $ {\cal A} $ be a bigraded
algebra with
bidegree called {\em resolution bidegree}. We set
\begin{equation}
\label{bires}
bires = (res_{1},res_{2})
\end{equation}
and
\begin{equation}
\label{res}
res = res_{1} + res_{2}.
\end{equation}
We assume that both $ res_{1} $ and $ res_{2} $  are non negative
integers :
$ res_{1} \geq 0 $ and $ res_{2} \geq 0 $.
\begin{biresolution}
\label{biresolutiondef}
Let $ \delta : {\cal A} \rightarrow {\cal A} $ be a differential of
resolution degree $ -1 $,
\begin{equation}
\label{d2=0}
\delta^{2} = 0,
\end{equation}
\begin{equation}
\label{redd=-1}
res(\delta) = -1,
\end{equation}
i.e.
\begin{eqnarray}
res(\delta a) & = & res(a) - 1 \qquad {\it when} \qquad res(a)  \geq
1,
\label{res(da)1} \\
& = & 0  \qquad {\it when} \qquad  res(a)  =  0, \ ({\it in \ which \
case} \
\delta a = 0). \label{res(da)2}
\end{eqnarray}
One says that the differential complex $ ({\cal A},\delta) $ is a
biresolution
of the algebra $ {\cal A}_{0} $ if and only if:
\begin{enumerate}
\item The differential $ \delta $ splits as the sum of two derivations
only
\begin{equation}
\label{d=d1+d2}
\delta = \delta_{1} + \delta_{2}
\end{equation}
with
\begin{equation}
\label{biresd1d2}
bires(\delta_{1}) = (-1,0), \qquad bires(\delta_{2}) = (0,-1)
\end{equation}
(no extra piece, say, of resolution bidegree $ (-2,1) $). It follows
from the
nilpotency of $ \delta $ that
\begin{equation}
\delta_{1}^{2} = \delta_{1}\delta_{2} + \delta_{2}\delta_{1} =
\delta_{2}^{2} = 0.
\end{equation}
i.e. $ \delta_{1}$ and $ \delta_{2} $ are differentials that
anticommute.
\item One has
\begin{eqnarray}
H_{0,0}(\delta_{1}) & = & {\cal A}_{0}, \  H_{0,k}(\delta_{1}) = 0 =
H_{k,*}(\delta_{1}), \qquad (k \not= 0) \label{coh1} \\
H_{0,0}(\delta_{2}) & = & {\cal A}_{0}, \  H_{k,0}(\delta_{2}) = 0 =
H_{*,k}(\delta_{2}), \qquad (k \not= 0) \label{coh2} \\
H_{0}(\delta) & = & {\cal A}_{0}, \  H_{k}(\delta) = 0, \qquad
(k \not= 0). \label{coh3}
\end{eqnarray}
\end{enumerate}
\end{biresolution}

\noindent
{\em Remark} : the relation (\ref{coh3}) is easily seen to be a
consequence
of (\ref{coh1}) and (\ref{coh2}).

\newtheorem{SymBiRes}[biresolution]{Definition}
\begin{SymBiRes}
\label{SymBiRes}
A biresolution is said to be symmetric if there exists an involution $
S $
($ S^{2} = 1 $) which (i) is an algebra isomorphism ; (ii) maps an
element
of bidegree $ (a,b) $ on an element of bidegree $ (b,a) $ and (iii)
maps
$ \delta_{1} $ on $ \delta_{2} $ and vice-versa :
\begin{eqnarray}
S \delta_{1} S & = & \delta_{2} \label{inv1}, \\
S \delta_{2} S & = & \delta_{1} \label{inv2}.
\end{eqnarray}
\end{SymBiRes}
Note that the relations (\ref{inv1}) and (\ref{inv2}) imply
\begin{equation}
\label{invd}
S \delta S = \delta.
\end{equation}

\subsection{Basic properties of biresolutions}
\label{Subsec33}
\newtheorem{Ther31}{Theorem}[section]
\begin{Ther31}
\label{Ther31}
Let $ ({\cal A},\delta) $ be a biresolution and
$ {\raise1pt\hbox{$ \stackrel{(a,b)}{F}$}} \in {\cal A} $,
$ bires({\raise1pt\hbox{$ \stackrel{(a,b)}{F}$}}) = (a,b) $
(with $ a + b > 0 $) be such
that
\begin{equation}
\left\{
\begin{array}{lcr}
\delta_{1} {\raise1pt\hbox{$ \stackrel{(a,b)}{F}$}} & = & 0 \\
\delta_{2} {\raise1pt\hbox{$ \stackrel{(a,b)}{F}$}} & = & 0
\end{array} \right. \qquad \Longleftrightarrow \qquad
\delta {\raise1pt\hbox{$ \stackrel{(a,b)}{F}$}} = 0.
\end{equation}
Then
\begin{equation}
{\raise1pt\hbox{$ \stackrel{(a,b)}{F}$}} = \delta_{2} \delta_{1}
{\raise1pt\hbox{$ \stackrel{(a+1,b+1)}{M}$}}.
\end{equation}
\end{Ther31}
{\bf Proof of theorem \ref{Ther31}}: From $ \delta_{2}
{\raise1pt\hbox{$ \stackrel{(a,b)}{F}$}} = 0 $, one gets
\begin{equation}
\label{eq1}
{\raise1pt\hbox{$ \stackrel{(a,b)}{F}$}} = \delta_{2}
{\raise1pt\hbox{$ \stackrel{(a,b+1)}{R}$}}
\end{equation}
since $ H_{a,b}(\delta_{2}) = 0 $ for $ a+b > 0 $. But one has also
$ \delta_{1} {\raise1pt\hbox{$ \stackrel{(a,b)}{F}$}} = 0 $, hence
$ \delta_{2} \delta_{1} {\raise1pt\hbox{$ \stackrel{(a,b+1)}{R}$}} = 0
$, i.e.,
there exists $ {\raise1pt\hbox{$ \stackrel{(a-1,b+2)}{R}$}} $ such that
\begin{equation}
\delta_{1} {\raise1pt\hbox{$ \stackrel{(a,b+1)}{R}$}} +
\delta_{2} {\raise1pt\hbox{$ \stackrel{(a-1,b+2)}{R}$}} = 0.
\end{equation}
This leads to the {\em descent equations}
\begin{equation}
\delta_{1} {\raise1pt\hbox{$ \stackrel{(a-1,b+2)}{R}$}} +
\delta_{2} {\raise1pt\hbox{$ \stackrel{(a-2,b+3)}{R}$}} = 0
\end{equation}
\begin{displaymath}
\vdots
\end{displaymath}
\begin{equation}
\label{desc2}
\delta_{1}{\raise1pt\hbox{$ \stackrel{(1,a+b)}{R}$}} +
\delta_{2}{\raise1pt\hbox{$ \stackrel{(0,a+b+1)}{R}$}} = 0
\end{equation}
\begin{equation}
\delta_{1}{\raise1pt\hbox{$ \stackrel{(0,a+b+1)}{R}$}} = 0.
\end{equation}
{}From the last equation and (\ref{coh1}), one obtains
\begin{equation}
{\raise1pt\hbox{$ \stackrel{(0,a+b+1)}{R}$}} = \delta_{1}
{\raise1pt\hbox{$ \stackrel{(1,a+b+1)}{M}$}}.
\end{equation}
Injecting this result in equation (\ref{desc2}), one gets
\begin{equation}
\delta_{1} \left( {\raise1pt\hbox{$ \stackrel{(1,a+b)}{R}$}} -
\delta_{2} {\raise1pt\hbox{$ \stackrel{(1,a+b+1)}{M}$}} \right) = 0,
\end{equation}
i.e., from (\ref{coh1})
\begin{equation}
{\raise1pt\hbox{$ \stackrel{(1,a+b)}{R}$}} =
\delta_{2} {\raise1pt\hbox{$ \stackrel{(1,a+b+1)}{M}$}} + \delta_{1}
{\raise1pt\hbox{$ \stackrel{(2,a+b)}{M}$}}.
\end{equation}
Going up the ladder in the same fashion, one finally gets for
$ {\raise1pt\hbox{$ \stackrel{(a,b+1)}{R}$}} $,
\begin{equation}
{\raise1pt\hbox{$ \stackrel{(a,b+1)}{R}$}} = \delta_{2}
{\raise1pt\hbox{$ \stackrel{(a,b+2)}{M}$}} + \delta_{1}
{\raise1pt\hbox{$ \stackrel{(a+1,b+1)}{M}$}}
\end{equation}
and thus, from (\ref{eq1})
\begin{equation}
{\raise1pt\hbox{$ \stackrel{(a,b)}{F}$}} = \delta_{2} \delta_{1}
{\raise1pt\hbox{$ \stackrel{(a+1,b+1)}{M}$}}.
\end{equation}
{\bf QED}

\newtheorem{Ther32}[Ther31]{Theorem}
\begin{Ther32}
\label{Ther32}
Let $ {\raise1pt\hbox{$ \stackrel{(m)}{F}$}} \in {\cal A} $, with $
res(
{\raise1pt\hbox{$ \stackrel{(m)}{F}$}}) = m > 0 $, be such that
\begin{equation}
\label{Ther321}
{\raise1pt\hbox{$ \stackrel{(m)}{F}$}} = \sum_{p+q=m}
{\raise1pt\hbox{$ \stackrel{(p,q)}{F}$}}.
\end{equation}
Assume that :
\begin{enumerate}
\item $ \delta {\raise1pt\hbox{$ \stackrel{(m)}{F}$}} = 0 $,
\item In the sum (\ref{Ther321}), only terms with $ p \leq a $ and $ q
\leq b $ occur,
for some $      a $ and $ b $ such that $ a + b > m $ (strictly).
\end{enumerate}
Then,
\begin{equation}
\label{Ther322}
{\raise1pt\hbox{$ \stackrel{(m)}{F}$}} = \delta
{\raise1pt\hbox{$ \stackrel{(m+1)}{P}$}}
\end{equation}
where
\begin{equation}
\label{Ther323}
{\raise1pt\hbox{$ \stackrel{(m+1)}{P}$}} = \sum_{\bar{p} + \bar{q} =
m+1}
{\raise1pt\hbox{$ \stackrel{(\bar{p},\bar{q})}{P}$}}
\end{equation}
involves only terms $ {\raise1pt\hbox{$
\stackrel{(\bar{p},\bar{q})}{P}$}} $
with $ \bar{p} \leq a $ and $ \bar{q} \leq b $.
\end{Ther32}
{\bf Proof of theorem \ref{Ther32}}: One has
\begin{equation}
\label{Ther324}
{\raise1pt\hbox{$ \stackrel{(m)}{F}$}} =
{\raise1pt\hbox{$ \stackrel{(a,m-a)}{F}$}} + \cdots +
{\raise1pt\hbox{$ \stackrel{(m-b,b)}{F}$}}
\end{equation}
(with $ {\raise1pt\hbox{$ \stackrel{(i,j)}{F}$}} = 0 $ if $ i < 0 $ or
$ j < 0 $).
{}From $ \delta {\raise1pt\hbox{$ \stackrel{(m)}{F}$}} = 0 $, one gets
$ \delta_{2} {\raise1pt\hbox{$ \stackrel{(a,m-a)}{F}$}} = 0 $, i.e.,
using
(\ref{coh2}),
\begin{equation}
\label{Ther325}
{\raise1pt\hbox{$ \stackrel{(a,m-a)}{F}$}} = \delta_{2}
{\raise1pt\hbox{$ \stackrel{(a,m-a+1)}{P'}$}}
\end{equation}
(if $ m-a < 0 $, ${\raise1pt\hbox{$ \stackrel{(a,m-a)}{F}$}} = 0 $ and
one takes
$ {\raise1pt\hbox{$ \stackrel{(a,m-a+1)}{P'}$}} \equiv 0 $). One has
$ m-a+1 \leq b $ because $ m < a+b $. If one substracts
$ \delta {\raise1pt\hbox{$ \stackrel{(a,m-a+1)}{P'}$}} $ from
$ {\raise1pt\hbox{$ \stackrel{(m)}{F}$}} $, one obtains
\begin{equation}
\label{Ther326}
{\raise1pt\hbox{$ \stackrel{(m)}{F}$}} - \delta
{\raise1pt\hbox{$ \stackrel{(a,m-a+1)}{P'}$}} =
{\raise1pt\hbox{$ \stackrel{(a-1,m-a+1)}{F'}$}} + \cdots +
{\raise1pt\hbox{$ \stackrel{(m-b,b)}{F}$}}.
\end{equation}
One then keeps going (one removes
$ {\raise1pt\hbox{$ \stackrel{(a-1,m-a+1)}{F'}$}} $,etc,$\ldots$) until
one reaches the
last step,
\begin{eqnarray}
\label{Ther327}
{\raise1pt\hbox{$ \stackrel{(m)}{F}$}} - \delta \tilde{P} & = &
{\raise1pt\hbox{$ \stackrel{(m-b,b)}{F'}$}} \\
\delta {\raise1pt\hbox{$ \stackrel{(m-b,m)}{F'}$}} & = & 0
\Longleftrightarrow
\left\{
\begin{array}{lll}
\delta_{1} {\raise1pt\hbox{$ \stackrel{(m-b,m)}{F'}$}} & = & 0 \\
\delta_{2} {\raise1pt\hbox{$ \stackrel{(m-b,m)}{F'}$}} & = & 0.
\end{array}
\right.
\end{eqnarray}
{}From theorem \ref{Ther31}, this implies that
\begin{eqnarray}
\label{Ther328}
{\raise1pt\hbox{$ \stackrel{(m-b,m)}{F'}$}} & = & \delta_{1} \delta_{2}
{\raise1pt\hbox{$ \stackrel{(m-b+1,m+1)}{S}$}} \nonumber \\
& = & (\delta_{1} + \delta_{2}) \delta_{2}
{\raise1pt\hbox{$ \stackrel{(m-b+1,m+1)}{S}$}} \nonumber \\
& = & \delta {\raise1pt\hbox{$ \stackrel{(m-b+1,m)}{Q}$}}
\end{eqnarray}
with $ {\raise1pt\hbox{$ \stackrel{(m-b+1,m)}{Q}$}} = \delta_{2}
{\raise1pt\hbox{$ \stackrel{(m-b+1,m+1)}{S}$}} $. One has $ m-b+1 \leq
a $ because
$ m < a + b $. {\bf QED}

\pagebreak

\noindent
\newtheorem{Ther33}[Ther31]{Theorem}
\begin{Ther33}
\label{Ther33}
Assume that in theorem \ref{Ther32}, $ {\raise1pt\hbox{$
\stackrel{(m)}{F}$}} $
is $ S $-even, i.e.,
\begin{equation}
\label{Ther331}
S{\raise1pt\hbox{$ \stackrel{(m)}{F}$}} = {\raise1pt\hbox{$
\stackrel{(m)}{F}$}}.
\end{equation}
Then $ {\raise1pt\hbox{$ \stackrel{(m+1)}{P}$}} $ in (\ref{Ther322})
can also
be chosen to be $ S $-even :
\begin{equation}
\label{Ther332}
S {\raise1pt\hbox{$ \stackrel{(m+1)}{P}$}} =
{\raise1pt\hbox{$ \stackrel{(m+1)}{P}$}}.
\end{equation}
Similarly, if $ {\raise1pt\hbox{$ \stackrel{(m)}{F}$}} $ is $ S $-odd,
$ {\raise1pt\hbox{$ \stackrel{(m+1)}{P}$}} $ can be chosen to be $ S $-
odd.
\end{Ther33}
{\bf Proof of theorem \ref{Ther33}}: We treat only the case
$ {\raise1pt\hbox{$ \stackrel{(m)}{F}$}} $ $S$-even. The case
$ {\raise1pt\hbox{$ \stackrel{(m)}{F}$}} $ $S$-odd is treated in a
similar
fashion. Because $ {\raise1pt\hbox{$ \stackrel{(m)}{F}$}} $ is $S$-
even, one
can assume $ a=b $ in the previous theorem. Now, from (\ref{Ther331}),
(\ref{Ther322}) and (\ref{invd}), one gets
\begin{equation}
\label{Ther333}
{\raise1pt\hbox{$ \stackrel{(m)}{F}$}} = \delta \left[
\frac{1}{2} \left( {\raise1pt\hbox{$ \stackrel{(m+1)}{P}$}} +
S {\raise1pt\hbox{$ \stackrel{(m+1)}{P}$}} \right) \right] .
\end{equation}
Both $ {\raise1pt\hbox{$ \stackrel{(m+1)}{P}$}} $ and
$ S {\raise1pt\hbox{$ \stackrel{(m+1)}{P}$}} $ fulfill the conditions
of theorem
\ref{Ther32} since $ a=b $. Clearly, $ 1 / 2 (
{\raise1pt\hbox{$ \stackrel{(m+1)}{P}$}} +
S {\raise1pt\hbox{$ \stackrel{(m+1)}{P}$}}) $ is $S$-even. {\bf QED}

\section{Koszul-Tate biresolution}
\label{Sec4}
\subsection{Koszul-Tate resolution}
To warm up, we shall first recall a standard result on Koszul-Tate
resolutions,
which has been derived in the context of BRST theory
\cite{Fish:Henneaux:Stasheff:Teitelboim,Henneaux:Teitelboim}. To that
end, we
come back to the geometrical data of section \ref{Subsec21}. The
equations
(\ref{tang}) defining the submanifold $ \Sigma \subset \Gamma $,
\begin{equation}
\label{constraints}
G_{A_{0}} \approx 0
\end{equation}
may not be independent, i.e., there may be relations among the $
G_{A_{0}} $'s :
\begin{equation}
\label{red1}
Z^{A_{0}}_{A_{1}} G_{A_{0}} = 0 \qquad ({\rm identically}).
\end{equation}
The functions $ Z^{A_{1}}_{A_{0}} $ are called the first order
reducibility
functions and provide a complete set of relations among the
constraints. They
may, in turn, be non independent, i.e., there may be relations among
them
\begin{equation}
\label{red2}
Z^{A_{1}}_{A_{2}} Z^{A_{1}}_{A_{0}} \approx 0,
\end{equation}
etc. There is thus a tower of reducibility identities of the form
\begin{equation}
\label{redk}
Z^{A_{k-1}}_{A_{k}} Z^{A_{k-2}}_{A_{k-1}} \approx 0,
\end{equation}
the last one being
\begin{equation}
\label{redL}
Z^{A_{L-1}}_{A_{L}} Z^{A_{L-2}}_{A_{L-1}} \approx 0.
\end{equation}

\newtheorem{complete set}{Definition}[section]
\begin{complete set}
\cite{Fish:Henneaux:Stasheff:Teitelboim,Henneaux:Teitelboim}
The set $ \{G_{A_{0}},Z^{A_{0}}_{A_{1}}, \ldots , Z^{A_{L-1}}_{A_{L}}\}
$
provides a complete description of $ \Sigma $ if $ z^{I}
\in \Sigma \Leftrightarrow
G_{A_{0}}(z^{I}) = 0 $, and if
\begin{equation}
\xi^{A_{0}}G_{A_{0}} = 0 \: \Leftrightarrow \: \xi^{A_{0}} =
\xi^{A_{1}}Z^{A_{0}}_{A_{1}} + \nu^{A_{0}B_{0}}G_{B_{0}},
\end{equation}
\begin{displaymath}
\vdots
\end{displaymath}
\begin{equation}
\xi^{A_{k}}Z^{A_{k-1}}_{A_{k}} \approx 0 \: \Leftrightarrow \:
\xi^{A_{k}} \approx
\xi^{A_{k+1}}Z^{A_{k}}_{A_{k+1}}.
\end{equation}
\begin{displaymath}
\vdots
\end{displaymath}
\begin{equation}
\xi^{A_{L}}Z^{A_{L-1}}_{A_{L}} \approx 0 \: \Leftrightarrow \:
\xi^{A_{L}} \approx
0,
\end{equation}
\vskip5truemm
where $ \nu^{A_{0}B_{0}} = (-
)^{(\epsilon_{A_{0}}+1)(\epsilon_{B_{0}}+1)}\nu^{B_{0}A_{0}} $.
\end{complete set}

\newtheorem{Koszul-Tate}{Theorem}[section]
\begin{Koszul-Tate}
\label{KosTate}
\cite{Fish:Henneaux:Stasheff:Teitelboim,Henneaux:Teitelboim}
To each complete description $ \{G_{A_{0}},Z^{A_{0}}_{A_{1}}, \ldots
,Z^{A_{L-1}}_{A_{L}}\} $
of the surface $\Sigma \subset \Gamma $, one can associate a graded
differential
complex $ (K_{*},\delta) $ such that
\begin{enumerate}
\item $ K = \mbox{{\helv C}}[{\cal P}_{A_{0}}, \ldots , {\cal
P}_{A_{L}}]
\otimes C^{\infty}(\Gamma) $, where
\footnote{Among the $ {\cal P}_{A_{0}}, \ldots , {\cal P}_{A_{L}} $,
some are
commuting and some are anticommuting (see
\cite{Fish:Henneaux:Stasheff:Teitelboim,Henneaux:Teitelboim}). We
denote by
$ \mbox{{\helv C}}[{\cal P}_{A_{0}}, \ldots , {\cal P}_{A_{L}}] $ the
algebra
of polynomials in these variables with complex coefficients. For
instance,
for $ \theta $ anticommuting,
$ \mbox{{\helv C}}[\theta] = \{ \alpha + \beta \theta \}, \alpha, \beta
\in
\mbox{{\helv C}} $. Although we allow complex coefficients, the concept
of
{\em smoothness} is used in the real sense. } $ res({\cal P}_{A_{n}}) =
n+1 $.
\item The operator $ \delta $ is defined on the generators of the
algebra
$ K $ by
\begin{eqnarray}
\delta z^{I} & = & 0 ,  \label{delta0} \\
\delta {\cal P}_{A_{0}} & = & - G_{A_{0}}, \label{delta1} \\
\delta{\cal P}_{A_{1}} & & = - Z^{A_{0}}_{A_{1}}{\cal P}_{A_{0}},
\label{delta2} \\
& \vdots & \nonumber \\
\delta{\cal P}_{A_{k}} & = & - Z^{A_{k-1}}_{A_{k}}{\cal P}_{A_{k-1}}
+ M_{A_{k}}[{\cal P}_{A_{0}},\ldots,{\cal P}_{A_{k-2}}], \label{delta3}
\\
& \vdots & \nonumber \\
\delta{\cal P}_{A_{L}} & = & - Z^{A_{L-1}}_{A_{L}}{\cal P}_{A_{L-1}}
+ M_{A_{L}}[{\cal P}_{A_{0}},\ldots,{\cal P}_{A_{L-2}}], \label{delta4}
\end{eqnarray}
where the functions $ M_{A_{k}} $ are such that the Koszul-Tate
operator $ \delta $ is nilpotent, $ \delta^{2} = 0 $.
\item $ H_{k}(\delta) = 0 $ for $ k>0 $ and $
H_{0}(\delta) = C^{\infty}(\Sigma) $ , that is, $ (K_{*},\delta) $
provides a
homological resolution of the algebra $ C^{\infty}(\Sigma) $.
\end{enumerate}
\end{Koszul-Tate}
The graded differential complex $ (K_{*},\delta) $ is the Koszul-Tate
differential complex and the associated resolution of $
C^{\infty}(\Sigma) $ is
called the Koszul-Tate resolution.
Conversely, if a differential of the form (\ref{delta0})-(\ref{delta4})
provides a homological resolution of $ C^{\infty}(\Sigma) $, then, the
functions
$ \{G_{A_{0}},Z^{A_{0}}_{A_{1}}, \ldots , Z^{A_{L-1}}_{A_{L}}\} $
appearing in
(\ref{delta0})-(\ref{delta4}) constitute a complete description of $
\Sigma $.

Our purpose in this section is to show that for each complete
description of the
constraint surface, one can also associate a Koszul-Tate biresolution,
by
repeating an appropriate number of times the constraints and the
reducibility
functions.

\subsection{Results}
\label{Subsec42}
We have indicated in \cite{Gregoire:Henneaux} the way in which one
should
proceed when the functions $ G_{A_{0}} $ defining $ \Sigma $ are
independent
(irreducible case). Rather than the single ``ghost momentum" $ {\cal
P}_{A_{0}} $
of resolution degree one, one should introduce two ghosts momenta
$ \ghost{.5}{1}{0}{\cal P}{A}{0} $ and $ \ghost{.5}{0}{1}{\cal P}{A}{0}
$ at
respective resolution bidegree $ (1,0) $ and $ (0,1) $. That is, one
duplicates
the constraints $ G_{A_{0}} \approx 0 $ by simply repeating them a
second time.
The description of $ \Sigma $ by means of the duplicated constraints is
clearly
no longer irreducible. One then introduces a ghost momentum
$ \ghost{0}{1}{1}{\lambda}{A}{0} $ to compensate for the duplication
and sets
\begin{eqnarray}
\delta \ghost{.5}{1}{0}{\cal P}{A}{0} & = & - G_{A_{0}}, \\
\delta \ghost{.5}{0}{1}{\cal P}{A}{0} & = & - G_{A_{0}}, \\
\delta \ghost{0}{1}{1}{\lambda}{A}{0} & = & \ghost{.5}{0}{1}{\cal
P}{A}{0} -
\ghost{.5}{1}{0}{\cal P}{A}{0}.
\end{eqnarray}
This defines the searched-for biresolution in the irreducible
hamiltonian
case. That biresolution is symmetric under the involution
\begin{eqnarray}
S \ghost{.5}{1}{0}{\cal P}{A}{0} & = & \ghost{.5}{0}{1}{\cal P}{A}{0},
\qquad
S \ghost{.5}{0}{1}{\cal P}{A}{0} = \ghost{.5}{1}{0}{\cal P}{A}{0}, \\
S \ghost{0}{1}{1}{\lambda}{A}{0} & = & -
\ghost{0}{1}{1}{\lambda}{A}{0}.
\end{eqnarray}

In the irreducible case, there are higher order ghost momenta $ {\cal
P}_{A_{k}} $
in the Koszul-Tate resolution, of resolution degree $ k+1 $. These
should be
replaced by $ (k+2) $ ghost of ghost momenta $ \ghost{.5}{i}{j}{\cal
P}{A}{k} $,
with $ i+j = k+1 $, $ i \geq 0 $, $ j \geq 0 $. This provides a
spectrum symmetric
for the interchange of $ i $ with $ j $. This also amounts to repeating
the
reducibility functions $ k+2 $ times, increasing thereby the
reducibility. One
thus needs further ghosts of ghost momenta $
\ghost{0}{i+1}{j+1}{\lambda}{A}{k} $,
with $ i+j = k $, $ i \geq 0 $, $ j \geq 0 $, in order to compensate
for that increase
in reducibility.

Rather then trying to give a systematic, step-by-step derivation of the
corresponding Koszul-Tate biresolution, we shall first state the
results and
then prove their correctness.

\newtheorem{SymBiresTheo}[Koszul-Tate]{Theorem}
\begin{SymBiresTheo}
\label{Ther42}
To each complete description
$ \{G_{A_{0}},Z^{A_{0}}_{A_{1}}, \ldots ,Z^{A_{L-1}}_{A_{L}}\} $ of the
constraint surface $ \Sigma \subset \Gamma $, one can associate a
symmetric
biresolution $ (K_{*}, \delta = \delta_{1} + \delta_{2}) $ of
$ C^{\infty}(\Sigma) $
defined as follows.
\begin{enumerate}
\item The graded algebra $ K_{*} $ is defined by
\begin{eqnarray}
K_{*} & = & \mbox{{\helv C}}[\ghost{.5}{i_{0}}{j_{0}}{\cal P}{A}{0} ,
\ghost{.5}{i_{1}}{j_{1}}{\cal P}{A}{1} , \ldots ,
\ghost{.5}{i_{L}}{j_{L}}{\cal P}{A}{L} , \nonumber \\
& & \ghost{0}{i'_{0}+1}{j'_{0}+1}{\lambda}{A}{0} , \ldots ,
\ghost{0}{i'_{L}+1}{j'_{L}+1}{\lambda}{A}{L} ] \otimes
C^{\infty}(\Gamma),
\label{gradedalg}
\end{eqnarray}
with
\begin{eqnarray}
i_{k} + j_{k} & = & k + 1, \qquad i_{k} \geq 0, j_{k} \geq 0,
\label{bires1} \\
i'_{k} + j'_{k} & = & k + 1, \qquad i'_{k} \geq 0, j'_{k} \geq 0,
\label{bires2}
\end{eqnarray}
\begin{eqnarray}
bires(\ghost{.5}{i_{k}}{j_{k}}{\cal P}{A}{k} ) & = & (i_{k},j_{k}),
\label{bires3} \\
bires(\ghost{.5}{i'_{k}}{j'_{k}}{\cal P}{A}{k} ) & = & (i'_{k},j'_{k}),
\label{bires4}
\end{eqnarray}
\begin{eqnarray}
\epsilon(\ghost{.5}{i_{k}}{j_{k}}{\cal P}{A}{k} ) = \epsilon_{A_{k}} +
k + 1,
\label{bires5} \\
\epsilon(\ghost{0}{i'_{k}+1}{j'_{k}+1}{\lambda}{A}{k} ) =
\epsilon_{A_{k}} + k
\label{bires6}
\end{eqnarray}
(where $ \epsilon_{A_{k}} $ is defined recursively through $
\epsilon(G_{A_{0}})=
\epsilon_{A_{0}}, \epsilon(Z_{A_{k}}^{A_{k-1}}) =
\epsilon_{A_{k}} + \epsilon_{A_{k-1}} $).
\item The operator $ \delta = \delta_{1} + \delta_{2} $ acts on the
generators
as
\begin{equation}
\label{defd1}
\left\{
\begin{array}{lll}
\delta_{1} z^{I} & = & 0, \\
\delta_{2} z^{I} & = & 0,
\end{array} \right.
\end{equation}
\begin{equation}
\label{defd2}
\left\{
\begin{array}{lll}
\delta_{1} \ghost{.5}{k}{0}{\cal P}{A}{k-1} & = & - Z^{A_{k-2}}_{A_{k-
1}}
\ghost{.5}{k-1}{0}{\cal P}{A}{k-2} + \ghost{0}{k-1}{0}{M}{A}{k-1} \\
\delta_{2} \ghost{.5}{k}{0}{\cal P}{A}{k-1} & = & 0
\end{array} \right. \qquad (k \geq 0),
\end{equation}
\begin{equation}
\label{defd3}
\left\{
\begin{array}{lll}
\delta_{1} \ghost{.5}{0}{k}{\cal P}{A}{k-1} & = & 0 \\
\delta_{2} \ghost{.5}{0}{k}{\cal P}{A}{k-1} & = & - Z^{A_{k-2}}_{A_{k-
2}}
\ghost{.5}{0}{k-1}{\cal P}{A}{k-2} + \ghost{0}{0}{k-1}{\bar M}{A}{k-1}
\end{array} \right. \qquad (k \geq 0),
\end{equation}
\begin{equation}
\label{defd4}
\left\{
\begin{array}{lll}
\delta_{1} \ghost{.5}{i}{j}{\cal P}{A}{k-1} & = & - \frac{1}{2}
Z^{A_{k-2}}_{A_{k-1}} \ghost{.5}{i-1}{j}{\cal P}{A}{k-2} +
\ghost{0}{i-1}{j}{M}{A}{k-1} \\
\delta_{2} \ghost{.5}{i}{j}{\cal P}{A}{k-1} & = & - \frac{1}{2}
Z^{A_{k-2}}_{A_{k-1}} \ghost{.5}{i}{j-1}{\cal P}{A}{k-2} +
\ghost{0}{i}{j-1}{\bar M}{A}{k-1}
\end{array} \right. \qquad
\begin{array}{ll}
(i \neq 0 \neq j,\\
i+j = k \geq 0),
\end{array}
\end{equation}
\begin{equation}
\label{defd5}
\left\{
\begin{array}{ll}
\begin{array}{lll}
\delta_{1} \ghost{0}{i+1}{j+1}{\lambda}{A}{k-2} & = &
- \ghost{.5}{i}{j+1}{\cal P}{A}{k-2} + \frac{1}{2} Z^{A_{k-3}}_{A_{k-
2}}
\ghost{0}{i}{j+1}{\lambda}{A}{k-3} \\
& & + \ghost{0}{i}{j+1}{N}{A}{k-2}
\end{array}
& (i \neq 0, k \geq 2) \\
\begin{array}{lll}
\delta_{2} \ghost{0}{i+1}{j+1}{\lambda}{A}{k-2} & = &
- \ghost{.5}{i+1}{j}{\cal P}{A}{k-2} + \frac{1}{2} Z^{A_{k-3}}_{A_{k-
2}}
\ghost{0}{i+1}{j}{\lambda}{A}{k-3} \\
& & + \ghost{0}{i+1}{j}{\bar N}{A}{k-2}
\end{array}
& (j \neq 0, k \geq 2)
\end{array} \right.
\end{equation}
\begin{equation}
\label{defd6}
\left\{
\begin{array}{lll}
\delta_{1} \ghost{0}{1}{j+1}{\lambda}{A}{k-2} & = & -
\ghost{.5}{0}{j+1}{\cal P}{A}{k-2} + \ghost{0}{0}{j+1}{N}{A}{k-2} \\
\delta_{2} \ghost{0}{i+1}{0}{\lambda}{A}{k-2} & = & -
\ghost{.5}{i+1}{0}{\cal P}{A}{k-2} + \ghost{0}{i+1}{0}{\bar N}{A}{k-2}
\end{array} \right. \qquad (k \geq 2).
\end{equation}
\end{enumerate}
The functions $ M_{A_{k-1}}, {\bar M}_{A_{k-1}}, N_{A_{k-2}},
{\bar N}_{A_{k-2}} $ depend only on $ {\cal P}_{A_{u}} $ with $ u \leq
k - 3 $ and
$ \lambda_{A_{s}} $ with $ s \leq k-4$. They are determined recursively
in such a way
that $ \delta^{2} = 0 $ (see below), and are such that
\begin{eqnarray}
S \ghost{0}{i}{j}{M}{A}{k-1} & = & \ghost{0}{j}{i}{\bar M}{A}{k-1}, \\
S \ghost{0}{i}{j}{N}{A}{k-2} & = & \ghost{0}{j}{i}{\bar N}{A}{k-2}
\end{eqnarray}
where $ S $ is the symmetry
\begin{eqnarray}
S \ghost{.5}{i}{j}{\cal P}{A}{k-1} & = & \ghost{.5}{j}{i}{\cal P}{A}{k-
1}, \\
S \ghost{0}{i+1}{j+1}{\lambda}{A}{k-2} & = &
- \ghost{0}{j+1}{i+1}{\lambda}{A}{k-2}.
\end{eqnarray}
\end{SymBiresTheo}

\subsection{Proof of theorem 4.2}
\label{Subsec43}
We define
\begin{equation}
K_{k} = \mbox{{\helv C}}[{\cal P}_{A_{0}}, \cdots, {\cal P}_{A_{k}},
\lambda_{A_{0}}, \cdots , \lambda_{A_{k-1}}] \otimes C^{\infty}{\Gamma}
\end{equation}
and observe that $ K_{L+1} = K_{*} $. The proof of theorem \ref{Ther42}
proceeds in steps.

\noindent
{\bf Step 1}:Assume that one has been able to find $ M_{A_{i}} $,
$ {\bar M}_{A_{i}} $ up to $ i = k -1 $ and $ N_{A_{i}} $, $ {\bar
N}_{A_{i}} $
up to $ i = k -2 $, in such a way that ({\em i}) $ \delta^{2} = 0 $ on
$ K_{k} $; ({\em ii}) $ \delta $ contains only pieces of bidegree
$ (-1,0) $ and $ (0,-1) $, that is, $ \delta = \delta_{1} + \delta_{2}
$; and
({\em iii}) $ S \delta S = \delta $. Then, it is easy to see that if
the
element $ a \in K_{i} $ with $ i < k $ fulfills both $ res(a) > 0 $ and
$ \delta_{\mu} a = 0 $, then $ a = \delta_{\mu} b $ with
$ b \in K_{i+1} $. If $ K_{i} = K_{L+1} = K_{k} $, then $ a =
\delta_{\mu} b $
with $ b \in K_{L+1} $. Here $ \delta_{\mu} $ stands for either
$ \delta_{1}, \delta_{2} $ or $ \delta $.

\noindent
{\bf Proof of step 1}:(a) We first consider the case $ \delta_{\mu} =
\delta_{1} $. Because $ \delta_{1} $ is $ C^{\infty}(\Gamma)$-linear,
one can
proceed locally on $ \Gamma $. Now, by a redefinition of the
constraints and of
the reducibility functions, one can assume that $ Z_{A_{j+1}}^{A_{j}}
Z_{A_{j}}^{A_{j-1}} = 0 $ (strongly and not just weakly), at least
locally.
In that case, the operator $ \delta_{1} $ takes the simple form
\begin{equation}
\left\{
\begin{array}{lll}
\delta_{1} z^{I} & = & 0 , \qquad \delta_{1} \ghost{.5}{1}{0}{\cal
P}{A}{0} =
- G_{A_{0}} \\
\delta_{1} \ghost{.5}{j}{0}{\cal P}{A}{j-1} & = & - Z_{A_{j-1}}^{A_{j-
2}}
\ghost{.5}{j-1}{0}{\cal P}{A}{j-2} \qquad (j \geq 2),
\end{array} \right.
\end{equation}
\begin{equation}
\left\{
\begin{array}{lll}
\delta_{1} \ghost{.5}{0}{j}{\cal P}{A}{j-1} & = & 0 \\
\delta_{1} \ghost{.5}{l}{m}{\cal P}{A}{j-1} & = & - \frac{1}{2}
Z^{A_{j-2}}_{A_{j-1}} \ghost{.5}{l-1}{m}{\cal P}{A}{j-2} \qquad
(l \neq 0 \neq m, l+m=j) \\
\delta_{1} \ghost{0}{l+1}{m+1}{\lambda}{A}{j-2} & = & -
\ghost{.5}{l}{m+1}{\cal P}{A}{j-2} + \frac{1}{2} Z^{A_{j-3}}_{A_{j-2}}
\ghost{.5}{l}{m+1}{\lambda}{A}{j-3} \qquad (l \neq 0, l+m=j-2) \\
\delta_{1} \ghost{0}{1}{m+1}{\lambda}{A}{j-2} & = & -
\ghost{.5}{0}{m+1}{\cal P}{A}{j-2}.
\end{array} \right.
\end{equation}
If one redefines the variables $ \ghost{.5}{l}{m}{\cal P}{A}{j-2} $, $
m \neq
0 $, as follows
\begin{eqnarray}
\ghost{0}{l}{m+1}{\mu}{A}{j-2} & = & -
\ghost{.5}{l}{m+1}{\cal P}{A}{j-2} + \frac{1}{2} Z^{A_{j-3}}_{A_{j-2}}
\ghost{.5}{l}{m+1}{\lambda}{A}{j-3} \qquad (l \neq 0, l+m=j-2) \\
\ghost{0}{0}{j-1}{\mu}{A}{j-2} & = & - \ghost{.5}{0}{j-1}{\cal P}{A}{j-
2},
\end{eqnarray}
one can rewrite $ \delta_{1} $ in the form
\begin{equation}
\label{AAA}
\left\{
\begin{array}{lll}
\delta_{1} z^{I} & = & 0 , \qquad \delta_{1} \ghost{.5}{1}{0}{\cal
P}{A}{0} =
- G_{A_{0}} \\
\delta_{1} \ghost{.5}{j}{0}{\cal P}{A}{j-1} & = & - Z_{A_{j-1}}^{A_{j-
2}}
\ghost{.5}{j-1}{0}{\cal P}{A}{j-2} \qquad (j \geq 2),
\end{array} \right.
\end{equation}
\begin{equation}
\left\{
\begin{array}{lll}
\delta_{1} \ghost{0}{l+1}{m+1}{\lambda}{A}{j-2} & = &
\ghost{0}{l}{m+1}{\mu}{A}{j-2} \\
\delta_{2} \ghost{0}{l}{m+1}{\mu}{A}{j-2} & = & 0
\end{array} \right. \qquad (l+m = j-2).
\end{equation}
Since $ \mu $ is the $ \delta_{1} $-variation of $ \lambda $, the
$ \lambda - \mu $ pairs cancel in $ \delta_{1} $-homology in $ K_{k} $,
except the unmatched variables $ \ghost{0}{l}{m}{\mu}{A}{k} $
(l+m = k+1), for which the corresponding $ \lambda $'s do not live in
$ K_{k} $ but in $ K_{k+1} $. Furthermore, since (\ref{AAA}) has the
standard
form of the resolution of $ C^{\infty}(\Sigma) $ given in theorem
\ref{KosTate} (with $ {\cal P}_{A_{j}} \rightarrow
\ghost{.5}{j+1}{0}{\cal P}{A}{j} $), the non trivial $ \delta_{1} $-
cycles
in $ K_{i} $ are all killed in $ K_{i+1} $ (or in $ K_{i} $ if $ i \geq
L $).
This proves step 1 for $ \delta_{1} $.

\noindent
(b) Step 1 for $ \delta_{2} $ is proved similarly.

\noindent
(c) The proof of step 1 for $ \delta = \delta_{1} + \delta_{2} $
follows
standard spectral sequence arguments. If $ \delta a = 0 $ with $ res(a)
= j > 0 $
and $ a \in K_{i} $, then $ a = \sum \ghost{0}{t}{j-t}{a}{}{} $. The
equation
$ \delta a = 0 $ implies $ \delta_{1} \ghost{0}{t_{min}}{j-
t_{min}}{a}{}{} = 0 $ for
the component of $ a $ with smallest $ t $. Then,
$ \ghost{0}{t_{min}}{j-t_{min}}{a}{}{} = \delta_{1}
\ghost{0}{t_{min}+1}{j-t_{min}}{b}{}{} $ with
$ \ghost{0}{t_{min}+1}{j-t_{min}}{b}{}{} \in K_{i+1} $ by (a), and the
component with smallest $ t $ of $ a - \delta
\ghost{0}{t_{min}+1}{j-t_{min}}{b}{}{} $ has $ t'_{min} = t_{min}+1 $.
Going on recursively along the same line, one easily arrives at the
desired
result.

\noindent
{\bf Step 2}: It is clear that if $ a \in K_{i} $ fulfills $ \delta a =
0 $,
$ res(a) > 0 $ and the positivity properties of theorem \ref{Ther32},
then
$ b \in K_{i+1} $ (or $ K_{L+1} $ if $ i = L+1 $) fulfills also the
positivity
properties of theorem \ref{Ther32}.

\noindent
{\bf Step 3}: $ \delta $ is defined on $ K_{0} $ and $ K_{1} $ by
\begin{eqnarray}
\delta z^{I} & = & 0, \\
\delta \ghost{.5}{1}{0}{\cal P}{A}{0} & = & - G_{A_{0}}, \qquad
\delta \ghost{.5}{0}{1}{\cal P}{A}{0} = - G_{A_{0}}, \\
\delta \ghost{.5}{2}{0}{\cal P}{A}{1} & = & Z^{A_{0}}_{A_{1}}
\ghost{.5}{1}{0}{\cal P}{A}{0}, \qquad
\delta \ghost{.5}{0}{2}{\cal P}{A}{1} = Z^{A_{0}}_{A_{1}}
\ghost{.5}{0}{1}{\cal P}{A}{0}, \\
\delta \ghost{.5}{1}{1}{\cal P}{A}{1} & = & - \frac{1}{2}
( \ghost{.5}{1}{0}{\cal P}{A}{0} + \ghost{.5}{0}{1}{\cal P}{A}{0} ), \\
\delta \ghost{0}{1}{1}{\lambda}{A}{0} & = & \ghost{.5}{0}{1}{\cal
P}{A}{0}
- \ghost{.5}{1}{0}{\cal P}{A}{0} .
\end{eqnarray}
It is such that $ \delta^{2} = 0 $, $ \delta = \delta_{1} + \delta_{2}
$ and
$ S \delta S =  \delta $. So let us assume that $ \delta $ has been
defined
on $ K_{i} $ up to $ i = k $, and let us show that one can extend $
\delta $
to $ K_{k+1} $, i.e., find $ M_{A_{k+1}} $, $ {\bar M}_{A_{k+1}} $,
$ N_{A_{k}} $ and $ {\bar N}_{A_{k}} \in K_{k-1} $ such that
$ \delta^{2} {\cal P}_{A_{k+1}} = \delta^{2} \lambda_{A_{k}} = 0 $,
$ \delta = \delta_{1} + \delta_{2} $ and
$ S \delta S {\cal P}_{A_{k+1}} = \delta {\cal P}_{A_{k+1}} $,
$ S \delta S \lambda_{A_{k}} = \delta \lambda_{A_{k}} $.
We shall only show how to define $ \delta \ghost{.5}{i}{j}{\cal
P}{A}{k+1} $ and
$ \delta \ghost{.5}{j}{i}{\cal P}{A}{k+1} $ $ (i>2, j>2) $, with $
i+j=k+2 $.
One proceeds along identical lines for the other variables.

Let $ M_{A_{k+1}} $ be the sum $ \ghost{0}{i-1}{j}{M}{A}{k+1} +
\ghost{0}{i}{j-1}{\bar M}{A}{k+1} $ and let $ {\bar M}_{A_{k+1}} $ be
$ \ghost{0}{j-1}{i}{M}{A}{k+1} + \ghost{0}{j}{i-1}{\bar M}{A}{k+1} $.
One
must find $ M_{A_{k+1}} $ and $ {\bar M}_{A_{k+1}} $ in $ K_{k-1} $
such that
the expressions
\begin{eqnarray}
\delta \ghost{.5}{i}{j}{\cal P}{A}{k+1} & = & - \frac{1}{2}
Z^{A_{k}}_{A_{k+1}} ( \ghost{.5}{i-1}{j}{\cal P}{A}{k} +
\ghost{.5}{i}{j-1}{\cal P}{A}{k} ) + M_{A_{k+1}}, \\
\delta \ghost{.5}{j}{i}{\cal P}{A}{k+1} & = & - \frac{1}{2}
Z^{A_{k}}_{A_{k+1}} ( \ghost{.5}{j-1}{i}{\cal P}{A}{k} +
\ghost{.5}{j}{i-1}{\cal P}{A}{k} ) + {\bar M}_{A_{k+1}}
\end{eqnarray}
have vanishing $ \delta $. Furthermore, $ M_{A_{k+1}} $ can contain
only terms
of bidegrees $ (i-1,j) $ and $ (i,j-1) $, and $ {\bar M}_{A_{k+1}} $
can contain
only terms of bidegree $ (j-1,i) $ and $ (j,i-1) $ (in order for $
\delta $ to
split as the sum of two differentials). Finally, one requires
$ S M_{A_{k+1}} = {\bar M}_{A_{k+1}} $.

Let $ X_{A_{k+1}} \in K_{k-1} $ be
\begin{eqnarray}
X_{A_{k+1}} & = & + (-)^{k} C^{A_{k-1}A_{0}}_{A_{k+1}} \left[
\frac{1}{4} \ghost{.5}{1}{0}{\cal P}{A}{0} \ghost{.5}{i-2}{j}{\cal
P}{A}{k-1}
\right. \nonumber \\
& & \left. + \frac{1}{4} (\ghost{.5}{1}{0}{\cal P}{A}{0}
+ \ghost{.5}{0}{1}{\cal P}{A}{0} )
\ghost{.5}{i-1}{j-1}{\cal P}{A}{k-1} + \frac{1}{4}
\ghost{.5}{0}{1}{\cal P}{A}{0} \ghost{.5}{i}{j-2}{\cal P}{A}{k-1}
\right]
\end{eqnarray}
where $ C^{A_{k-1}A_{0}}_{A_{k+1}} $ are the structure functions
appearing
in the identity
\begin{equation}
\label{structfunctions}
Z^{A_{k}}_{A_{k+1}} Z^{A_{k-1}}_{A_{k}} = (-)^{\epsilon_{A_{k-1}}}
C^{A_{k-1}A_{0}}_{A_{k+1}} G_{A_{0}}.
\end{equation}
Set
\begin{eqnarray}
M_{A_{k+1}} & = & M'_{A_{k+1}} + X_{A_{k+1}}, \\
{\bar M}_{A_{k+1}} & = & {\bar M}'_{A_{k+1}} + S X_{A_{k+1}}.
\end{eqnarray}
The unknown functions $ M'_{A_{k+1}} $ and $ {\bar M}'_{A_{k+1}} \in
K_{k-1} $
are subject to the equations :
\begin{eqnarray}
\delta M'_{A_{k+1}} = \delta D_{A_{k+1}}, \label{eqM'1} \\
\delta {\bar M}'_{A_{k+1}} =  \delta S D_{A_{k+1}} \label{eqM'2}
\end{eqnarray}
where
\begin{equation}
D_{A_{k+1}} = - X_{A_{k+1}} + \frac{1}{2} Z^{A_{k}}_{A_{k-1}}
(\ghost{.5}{i-1}{j}{\cal P}{A}{k} + \ghost{.5}{i}{j-1}{\cal P}{A}{k} )
\end{equation}
belongs to $ K_{k} $. Because $ \delta $ is nilpotent in (the already
constructed) $ K_{k} $, one has $ \delta ( \delta D_{A_{k+1}} ) = 0 $.
Furthermore, a straightforward calculation using identity
(\ref{structfunctions})
shows that $ \delta D_{A_{k+1}} \in K_{k-2} $. Hence, there exists
$ M'_{A_{k+1}} \in K_{k-1} $ such that the equation (\ref{eqM'1}) is
satisfied (see step 1). Note that $ M'_{A_{k+1}} \not= D_{A_{k+1}} $
because
$ D_{A_{k+1}} \in K_{k} $. Since $ \delta D_{A_{k+1}} $ contains
only terms of bidegrees
$ (i-2,j) $, $ (i-1,j-1) $ and $ (i,j-2) $, one infers,
using step 2 and theorem \ref{Ther32}, that $ M'_{A_{k+1}} $ can be
taken to
contain only terms of bidegrees $ (i-1,j) $ and $ (i,j-1) $, as
required.
Finally, one solves the second equation (\ref{eqM'2}) by taking
$ {\bar M}'_{A_{k+1}} = S M'_{A_{k+1}} $. This is acceptable because
$ S \delta = \delta S $ in the already constructed $ K_{k} $.

This completes the definition of $ \delta \ghost{.5}{i}{j}{\cal
P}{A}{k+1} $ and
$ \delta \ghost{.5}{j}{i}{\cal P}{A}{k+1} $. By a similar reasoning,
one defines
$ \delta $ on all the other new generators of $ K_{k+1} $, with the
required
properties. Step 3 Q and the proof of theorem \ref{Ther42} Q are
thereby
finished.

\section{BRST and anti-BRST generators}
\label{Sec5}
\subsection{Review of results from the BRST theory}
\label{Subsec51}
The existence of the Koszul-Tate biresolution is the hard core of the
BRST-anti-BRST theory. The rest of this paper merely takes advantage of
this
result by applying it in the context of standard BRST theory.

We recall that in the hamiltonian formulation of gauge theories, the
manifold
$ \Gamma $ is the phase space, with canonical coordinates $
(q^{i},p_{i}) $.
The functions $ G_{A_{0}} $ defining the constraint surface are first
class,
\begin{equation}
[G_{A_{0}},G_{B_{0}}] = C^{C_{0}}_{A_{0}B_{0}} G_{C_{0}}
\end{equation}
(after all the second class have been eliminated, e.g. through the
Dirac
bracket method). The observables are the equivalence classes of first
class
phase space functions $ F_{0} $ that coincide on the constraint surface
\begin{eqnarray}
[F_{0},G_{A_{0}}] & = & F^{B_{0}}_{A_{0}} G_{B_{0}}, \\
F_{0} & \sim & F_{0} + \lambda^{A_{0}} G_{A_{0}}.
\end{eqnarray}

One then has the important theorem
\cite{Stasheff,Fish:Henneaux:Stasheff:Teitelboim,DuboisViolette,Henneau
x:Teitelboim,Forger:Kellendonk}
\newtheorem{Ther51}{Theorem}[section]
\begin{Ther51}
\label{Ther51}
To any homological resolution $ (K_{*},\delta) $ of the constraint
surface, one can
associate a nilpotent function in an extended phase space:
\begin{equation}
\label{nileqomega}
[\Omega, \Omega] = 0, \qquad \epsilon(\Omega) = 1,
\end{equation}
which has the form
\begin{equation}
\label{Omegafterms}
\Omega = - \sum \eta ( \delta {\cal P}) + ``{\rm more}".
\end{equation}
The {\em BRST generator} $ \Omega $ generates the BRST transformation
through
\begin{equation}
s \cdot = [ \cdot, \Omega].
\end{equation}
The equation (\ref{nileqomega}) is equivalent to
\begin{equation}
s^{2} = 0
\end{equation}
and one has
\begin{equation}
H^{0}(s) \simeq C^{\infty}(\Sigma/{\cal G}) = \{ \ {\rm observables} \
\}.
\end{equation}
Actually, $ H^{*}(s) \simeq H^{*}(d) $ where $ d $ is the exterior
longitudinal
derivative along the gauge orbits on $ \Sigma $ for non negative degree
and
$ H^{*}(s) = 0 $ for negative degree.
\end{Ther51}

The variables $ \eta $ appearing in (\ref{Omegafterms}) are conjugate
to the
generators $ {\cal P} $ of the given resolution of $ C^{\infty}(\Sigma)
$. They
are called {\em ghosts}. The variables $ \eta^{A_{k}} $ with $ k > 0 $
are also
called {\em ghosts of ghosts}.

\subsection{Extended phase space}
The BRST-anti-BRST algebra (\ref{BRST-anti-BRSTdef}) implies
\begin{equation}
\label{s=s1+s2sq}
s^{2} \equiv (s_{1} + s_{2})^{2} = 0
\end{equation}
and conversely, (\ref{s=s1+s2sq}) implies (\ref{BRST-anti-BRSTdef})
provided
$ s $ splits as a sum of two differentials and no more
(if $ s $ were to split into more pieces, $s_{1}$ and $ s_{2} $ would
obey
equations involving the extra derivations contained in $ s $).
We shall use the previous theorem and the biresolution of section
\ref{Sec4} to
establish the existence of the BRST and anti-BRST generators. The idea
is the
same as that exposed in \cite{Gregoire:Henneaux} for the irreducible
case.
Namely, one constructs directly the generator $ \Omega $ of the sum
$ s_{1} + s_{2} $ by using the ordinary BRST theory, i.e. theorem
\ref{Ther51},
but applied to the description of $ \Sigma $ associated with the
differential
$ \delta $ of the previous section. And
one controls that $ \Omega $ splits as a sum of two terms only by means
of
theorem \ref{Ther32}.

The extended phase space is obtained by associated with each
$ \ghost{.5}{i}{j}{\cal P}{A}{k} $ and $
\ghost{0}{i+1}{j+1}{\lambda}{A}{k} $
of the previous section a conjugate ghost, denoted by $
\cghost{0}{i}{j}{\eta}{A}{k} $
or $ \cghost{0}{i+1}{j+1}{\pi}{A}{k} $~:
\begin{eqnarray}
[ \ghost{.5}{-i}{-j}{\cal P}{A}{k} , \cghost{0}{i'}{j'}{\eta}{A'}{k'} ]
& = &
\delta^{ii'} \delta^{jj'} \delta^{A'_{k'}}_{A_{k}} \label{eq113} \\
{[} \ghost{0}{-(i+1)}{-(j+1)}{\lambda}{A}{k} ,
\cghost{0}{i'+1}{j'+1}{\pi}{A'}{k'} ] & = & \delta^{ii'} \delta^{jj'}
\delta^{A'_{k'}}_{A_{k}}.
\end{eqnarray}
All the other brackets involving the ghosts or the ghost momenta
vanish.
The ghosts $ \cghost{0}{i}{j}{\eta}{A}{k} $ and
$ \cghost{0}{i+1}{j+1}{\pi}{A}{k} $ can be seen as the generators of a
model
for the longitudinal exterior differential complex $ (L^{*},d) $
\cite{Fish:Henneaux:Stasheff:Teitelboim,Henneaux:Teitelboim}. Actually,
this
model $ (K^{*},D) $ has a bicomplex structure and $ D = D_{1} + D_{2}
$.
The double differential complex $ (K^{*,*};D_{1},D_{2}) $ is bigraded
by the
{\em pure ghost bidegree}, denoted $ bipgh $ and defined by :
\begin{eqnarray}
bipgh(\cghost{0}{i}{j}{\eta}{A}{k} ) & = & (i,j) \\
bipgh(\cghost{0}{i+1}{j+1}{\pi}{A}{k} ) & = & (i+1,j+1).
\end{eqnarray}
The original canonical variables have zero $ bipgh $.
In so far as this does not play an important role in our construction,
we will
not elaborate more here on this aspect of the geometrical
interpretation of the
BRST-anti-BRST theory.

Following what is done in the usual BRST context, we also define
{\em ghost bidegree}, denoted $ bigh $ to be
\begin{equation}
bigh = bipgh - bires = (gh_{1},gh_{2})
\end{equation}
It is such that one has $ gh_{1}(s_{1}) = 1 = gh_{2}(s_{2}) $
and $ gh_{1}(s_{2}) = 0 = gh_{1}(s_{2}) $.
Also, one defines the {\em ghost degree} $ gh = gh_{1} + gh_{2} $.
{}From now on the superscript $ (i,j) $ will always indicates the ghost
bidegree.
So $ \ghost{.5}{i}{j}{\cal P}{A}{k} $ becomes
$ \ghost{.5}{-i}{-j}{\cal P}{A}{k} $ Qas already anticipated in
(\ref{eq113}).
We denote the bigraded polynomial algebra of polynomials in the ghosts
and the ghosts
momenta with coefficients that are functions of the original canonical
variables by $ {\cal K}^{*,*} $. One extends the definition of $ \delta
$ on
$ {\cal K}^{*,*} $ by requiring that $ \delta \eta = 0 = \delta \pi $;
with
this definition of $ \delta $, one has that $ bigh(\delta_{1}) = (1,0)
$ and
$ bigh(\delta_{2}) = (0,1) $. From the point of view of the BRST
theory based on $ \delta $, the variables $ \eta^{A_{k}} $ with $ k>0 $
and
$ \pi^{A_{k}} $ are the ghosts of ghosts associated with the reducible
description of $ \Sigma $ defined by $ \delta $. The degrees $ res $
and $ gh $
are respectively the corresponding resolution degree and ghost number.

\subsection{A positivity theorem}
\label{Subsec53}
\newtheorem{Positive poly}{Definition}[section]
\label{Positive poly}
\begin{Positive poly}
Let $ F \in {\cal K}^{*,*} $.If the polynomial $ F $ satisfied $
gh(F)=k>0 $
(respectively $ gh(F)=k \geq 0$), then $ F $ is said to be of
{\em positive ghost bidegree} (respectively {\em non negative ghost
bidegree})
if it can be decomposed as
\begin{equation}
F = \sum_{i+j=k} \raise1.1pt\hbox{ $ \stackrel{(i,j)}{F} $} ,
\end{equation}
where $ i \geq 0 $, $ j \geq 0 $ and $ bigh(\raise1.1pt\hbox{ $
\stackrel{(i,j)}{F} $})=(i,j) $. The
algebra of polynomials of positive ghost bidegree (respectively non
negative
ghost bidegree) is denoted by
$ {\cal K}^{*,*}_{++} $ (respectively ${\cal K}^{*,*}_{+} $). In
particular, one
has $ {\cal K}^{*,*}_{++} \subset {\cal K}^{*,*}_{+} $.
\end{Positive poly}

We have the following important theorem
\newtheorem{Ther52}[Ther51]{Theorem}
\begin{Ther52}
\label{Ther52}
Let $ F \in {\cal K}^{*,*}_{++} $ be such that (i) $res(F)=m>0 $ and
(ii)
$ \delta F = 0 $. Then, $ \exists P \in {\cal K}^{*,*}_{+} $ such that
$ \delta P = F $.
\end{Ther52}
{\bf Proof of theorem \ref{Ther52}} : Let $ F[\alpha,\beta;r,s] $
be the component of $ F $ satisfying $ bipgh(F[\alpha,\beta;r,s]) =
(\alpha,\beta) $
and $ bires(F[\alpha,\beta;r,s]) = (r,s) $. The condition $ \delta F =
0 $
implies $ \delta \sum_{r+s = k}F[\alpha,\beta;r,s] = 0 $, while the
condition
$ F \in {\cal K}^{*,*}_{+} $ implies $ r \leq \alpha $ and $ s \leq
\beta $ with
$ \alpha + \beta > m $. Applying theorem \ref{Ther32}, one obtains that
there exists $ P[\alpha,\beta] = \sum_{\bar{r} + \bar{s} = m+1}
P[\alpha,\beta;\bar{r},\bar{s}] $ such that $ {\bar r} \leq \alpha $
and
$ {\bar s} \leq \beta $. Thus $ P = \sum_{\alpha,\beta} P[\alpha,\beta]
\in {\cal K}^{*,*}_{+} $. {\bf QED}

\subsection{BRST and anti-BRST generators}
Let us consider the homological resolution $
\delta=\delta_{1}+\delta_{2} $
of theorem \ref{Ther42}. By theorem
\ref{Ther51}, we know that there exists a {\em total BRST charge} $
\Omega $,
that starts as
\begin{equation}
\Omega = G_{A_{0}}(\stackrel{(1,0)}{\eta^{A_{0}}}+
\stackrel{(0,1)}{\eta^{A_{0}}}) +
({\raise.5pt\hbox{$ \stackrel{(0,-1)}{{\cal P}_{A_{0}}}$}}-
{\raise.5pt\hbox{$ \stackrel{(-1,0)}{{\cal P}_{A_{0}}}$}})
\stackrel{(1,1)}{\pi^{A_{0}}} + \cdots \label{boundcond}
\end{equation}
However, we want more than just a mere solution of $ [\Omega,\Omega] =
0 $. We
want this solution to incorporate the full BRST-anti-BRST algebra. As
stressed already
above, this means that the total BRST transformation
$ s = [\cdot,\Omega^{T}] $ must split in two pieces $ s_{1} $ and $
s_{2} $
of different degrees. Accordingly, we require the total BRST generator
$ \Omega$ itself to split also in two pieces $ \Omega_{1} $ and $
\Omega_{2} $ with
$ bigh(\Omega_{1})=(1,0) $ and $ bigh(\Omega_{2})=(0,1) $. If this is
the case,
then the differentials $ s_{1} $ and $ s_{2} $ defined by
$ s_{1}=[\cdot,\Omega_{1}] $ and $ s_{2}=[\cdot,\Omega_{2}] $ fulfill
(\ref{BRST-anti-BRSTdef}).
\newtheorem{eqsplitting}[Ther51]{Theorem}
\begin{eqsplitting}
\label{eqsplitting}
Suppose that $ \Omega=\Omega_{1}+\Omega_{2} $ with
$ bigh(\Omega_{1})=(1,0) $ and $ bigh(\Omega_{2})$ $=(0,1) $, then
$ [\Omega,\Omega]=0 $ if and only if
$ [\Omega_{1},\Omega_{1}]=0=[\Omega_{2},\Omega_{2}] $ and
$ [\Omega_{1},\Omega_{2}]=0 $.
\end{eqsplitting}
{\bf Proof of theorem \ref{eqsplitting}}:Obvious by degree counting
arguments:
\begin{eqnarray}
[\Omega,\Omega] & = & [\Omega_{1} + \Omega_{2},
\Omega_{1} + \Omega_{2}] \nonumber \\
& = & [\Omega_{1},\Omega_{1}]+[\Omega_{2},\Omega_{2}]+
2[\Omega_{1},\Omega_{2}]. \label{split of omega}
\end{eqnarray}
Clearly, $ bigh([\Omega_{1},\Omega_{1}])=(2,0) $,
$ bigh([\Omega_{2},\Omega_{2}])=(0,1) $ and
$ bigh([\Omega_{1},\Omega_{2}])=(1,1) $. Thus, $ [\Omega,\Omega] =0$
if and only if each term of the right hand side of (\ref{split of
omega})
vanishes, that is, if and only if
$ [\Omega_{1},\Omega_{1}]=0=[\Omega_{2},\Omega_{2}] $ and
$ [\Omega_{1},\Omega_{2}]=0 $. {\bf QED}

We now prove that the total BRST charge can be split in just two pieces
$ \Omega_{1} $ and $ \Omega_{2} $.
\newtheorem{Splitting}[Ther51]{Theorem}
\begin{Splitting}
\label{Splitting}
One can choose the extra terms in (\ref{boundcond}) such that (i) $
\Omega $
splits as a sum of two terms of definite ghost bidegree, $
\Omega=\Omega_{1}+\Omega_{2} $
 with $ bigh(\Omega_{1})=(1,0) $ and $ bigh(\Omega_{2})=(0,1) $ and
(ii) $ [\Omega,\Omega] =0 $.
\end{Splitting}
{\bf Proof of theorem \ref{Splitting}}:Using homological perturbation
theory, the equation
$ [\Omega,\Omega] =0$ is equivalent to the family
\begin{equation}
\delta{\raise0pt\hbox{$ \stackrel{(n)}{\Omega}$}}=
{\raise1pt\hbox{$ \stackrel{(n-1)}{D}$}}[{\raise0pt\hbox{$
\stackrel{(0)}{\Omega}$}},
\ldots,{\raise0pt\hbox{$ \stackrel{(n-1)}{\Omega}$}}], \qquad n>0
\label{familyeq}
\end{equation}
where $ res({\raise0pt\hbox{$ \stackrel{(n)}{\Omega}$}})=n $.
The explicit form of
$ {\raise1pt\hbox{$ \stackrel{(n-1)}{D}$}} $, given in
\cite{Fish:Henneaux:Stasheff:Teitelboim},
is
\begin{eqnarray}
{\raise1pt\hbox{$ \stackrel{(n-1)}{D}$}}[{\raise0pt\hbox{$
\stackrel{(0)}{\Omega}$}},
& \ldots & ,{\raise0pt\hbox{$ \stackrel{(n-1)}{\Omega}$}}] =
{1 \over 2} \left\{ \sum_{m=0}^{n-1}[{\raise0pt\hbox{$ \stackrel{(n-m-
1)}{\Omega}$}},
{\raise0pt\hbox{$ \stackrel{(m)}{\Omega}$}}]_{orig} \right. \nonumber
\\
& + & \left. \sum_{m=1}^{n-1} \sum_{k=0}^{m-1} \left\{
[{\raise0pt\hbox{$ \stackrel{(n-m+k)}{\Omega}$}},
{\raise0pt\hbox{$ \stackrel{(m)}{\Omega}$}}]_{({\cal
P}_{A_{k}},\eta^{A_{k}})}
 + [{\raise0pt\hbox{$ \stackrel{(n+1-m+k)}{\Omega}$}},
{\raise0pt\hbox{$
\stackrel{(m)}{\Omega}$}}]_{(\lambda_{A_{k}},\pi^{A_{k}})}
\right\} \right\} , \nonumber \\
\  \label{ExplicitD}
\end{eqnarray}
where $ [\cdot,\cdot]_{orig} $ refers to the original Poisson bracket
not involving
the ghosts, $ [\cdot,\cdot]_{({\cal P}_{A_{k}},\eta^{A_{k}})}$ and
$ [\cdot,\cdot]_{(\lambda_{A_{k}},\pi^{A_{k}})}$
denote respectively the Poisson bracket with respect to the ghost pairs
$ ({\cal P}_{A_{k}},\eta^{A_{k}}) $ and
$ (\lambda_{A_{k}},\pi^{A_{k}}) $. Clearly, one
has $ {\raise0pt\hbox{$ \stackrel{(0)}{\Omega}$}}=
{\raise0pt\hbox{$ \stackrel{(0)}{\Omega_{1}}$}}+
{\raise0pt\hbox{$ \stackrel{(0)}{\Omega_{2}}$}} $. Suppose now that
$ {\raise0pt\hbox{$ \stackrel{(j)}{\Omega}$}}=
{\raise0pt\hbox{$ \stackrel{(j)}{\Omega_{1}}$}}+
{\raise0pt\hbox{$ \stackrel{(j)}{\Omega_{2}}$}} $ , for $ j<n $, then
let us
prove that $ {\raise0pt\hbox{$ \stackrel{(n)}{\Omega}$}} $ can be
chosen
such that $ {\raise0pt\hbox{$ \stackrel{(n)}{\Omega}$}}=
{\raise0pt\hbox{$ \stackrel{(n)}{\Omega_{1}}$}}+
{\raise0pt\hbox{$ \stackrel{(n)}{\Omega_{2}}$}} $ with
$ bigh(\Omega_{1})=(1,0) $ and $ bigh(\Omega_{2})=(0,1) $. Actually,
using
definition \ref{Positive poly}, one can reformulate this property as
follows.
Suppose that $ {\raise0pt\hbox{$ \stackrel{(j)}{\Omega}$}} $ is of
positive ghost bidegree for $ j <n $, then, we must show that
$ {\raise0pt\hbox{$ \stackrel{(n)}{\Omega}$}} $ may be chosen to be of
positive
ghost bidegree.
\newtheorem{PositiveD}[Ther51]{Lemma}
\begin{PositiveD}
\label{PositiveD}
Suppose that $ {\raise0pt\hbox{$ \stackrel{(j)}{\Omega}$}} $ is of
positive ghost bidegree for $ j<n $, then
$ {\raise1pt\hbox{$ \stackrel{(n-1)}{D}$}} $ is of positive ghost
bidegree.
\end{PositiveD}
{\bf Proof of lemma \ref{PositiveD}}: We observe that
$ {\raise1pt\hbox{$ \stackrel{(n-1)}{D}$}} $ is as follows :
\begin{eqnarray}
{\raise1pt\hbox{$ \stackrel{(n-1)}{D}$}}[{\raise0pt\hbox{$
\stackrel{(0)}{\Omega}$}},
\ldots,{\raise0pt\hbox{$ \stackrel{(n-1)}{\Omega}$}}] & = &
{\raise1pt\hbox{$ \stackrel{(n-1)}{D_{11}}$}}[{\raise0pt\hbox{$
\stackrel{(0)}{\Omega_{1}}$}},
\ldots,{\raise0pt\hbox{$ \stackrel{(n-1)}{\Omega_{1}}$}}] \nonumber \\
& + & {\raise1pt\hbox{$ \stackrel{(n-1)}{D_{22}}$}}[{\raise0pt\hbox{$
\stackrel{(0)}{\Omega_{2}}$}},
\ldots,{\raise0pt\hbox{$ \stackrel{(n-1)}{\Omega_{2}}$}}] \nonumber \\
& + & {\raise1pt\hbox{$ \stackrel{(n-1)}{D_{12}}$}}[{\raise0pt\hbox{$
\stackrel{(0)}{\Omega_{1}}$}},
\ldots,{\raise0pt\hbox{$ \stackrel{(n-1)}{\Omega_{1}}$}};
{\raise0pt\hbox{$ \stackrel{(0)}{\Omega_{2}}$}},
\ldots,{\raise0pt\hbox{$ \stackrel{(n-1)}{\Omega_{2}}$}}],
\end{eqnarray}
where $ {\raise0pt\hbox{$ \stackrel{(n-1)}{D_{11}}$}} $ stands for the
terms
computed from the sole $ {\raise0pt\hbox{$
\stackrel{(j)}{\Omega_{1}}$}}, j<n $,
$ {\raise0pt\hbox{$ \stackrel{(n-1)}{D_{22}}$}} $ from the sole
$ {\raise0pt\hbox{$ \stackrel{(j)}{\Omega_{2}}$}}, j<n $ and
$ {\raise0pt\hbox{$ \stackrel{(n-1)}{D_{12}}$}} $ for the mixed terms.
Using
(\ref{ExplicitD}), it is then easy to see that
\begin{equation}
bigh({\raise0pt\hbox{$ \stackrel{(n-1)}{D_{11}}$}})=(2,0),
\end{equation}
\begin{equation}
bigh({\raise0pt\hbox{$ \stackrel{(n-1)}{D_{22}}$}})=(0,2),
\end{equation}
\begin{equation}
bigh({\raise0pt\hbox{$ \stackrel{(n-1)}{D_{12}}$}})=(1,1).
\end{equation}
This clearly shows  that $ {\raise1pt\hbox{$ \stackrel{(n-1)}{D}$}} $
is of
positive ghost bidegree.  $ \triangleleft $

\noindent
Thus, in equation
$ \delta{\raise0pt\hbox{$ \stackrel{(n)}{\Omega}$}}=
{\raise1pt\hbox{$ \stackrel{(n-1)}{D}$}}[{\raise0pt\hbox{$
\stackrel{(0)}{\Omega}$}},
\ldots,{\raise0pt\hbox{$ \stackrel{(n-1)}{\Omega}$}}] $, the right hand
side
is of positive ghost bidegree and because
$ \delta{\raise1pt\hbox{$ \stackrel{(n-1)}{D}$}}=0 $, there
exists $ {\raise0pt\hbox{$ \stackrel{(n)}{\Omega}$}} $ such that ({\em
i})
$ \delta{\raise0pt\hbox{$ \stackrel{(n)}{\Omega}$}}=
{\raise0pt\hbox{$ \stackrel{(n-1)}{D}$}} $ and ({\em ii})
$ {\raise0pt\hbox{$ \stackrel{(n)}{\Omega}$}} $ is of positive ghost
bidegree (by theorem \ref{Ther52}). So, we have proven by induction
on the resolution degree that $ \Omega $ is of positive ghost bidegree
and this, in turn, implies that
\begin{equation}
\Omega=\Omega_{1}+\Omega_{2}
\end{equation}
with $ bigh(\Omega_{1})=(1,0) $ and $ bigh(\Omega_{2})=(0,1) $. {\bf
QED}

\pagebreak

A nice consequence of this theorem is that the family of equations
(\ref{familyeq})
can be decomposed in three pieces :
\begin{equation}
\label{nice1}
\delta_{1}{\raise0pt \hbox{$ \stackrel{(n)}{\Omega_{1}}$}}=
{\raise0pt\hbox{$ \stackrel{(n-1)}{D_{11}}$}}[{\raise0pt\hbox{$
\stackrel{(0)}{\Omega_{1}}$}},
\ldots,{\raise0pt\hbox{$ \stackrel{(n-1)}{\Omega_{1}}$}}],
\end{equation}
\begin{equation}
\label{nice2}
\delta_{2}{\raise0pt \hbox{$ \stackrel{(n)}{\Omega_{2}}$}}=
{\raise0pt\hbox{$ \stackrel{(n-1)}{D_{22}}$}}[{\raise0pt\hbox{$
\stackrel{(0)}{\Omega_{2}}$}},
\ldots,{\raise0pt\hbox{$ \stackrel{(n-1)}{\Omega_{2}}$}}],
\end{equation}
\begin{equation}
\label{nice3}
\delta_{1}{\raise0pt \hbox{$ \stackrel{(n)}{\Omega_{2}}$}}+
\delta_{2}{\raise0pt \hbox{$ \stackrel{(n)}{\Omega_{1}}$}}=
{\raise0pt\hbox{$ \stackrel{(n-1)}{D_{12}}$}}[{\raise0pt\hbox{$
\stackrel{(0)}{\Omega_{1}}$}},
\ldots,{\raise0pt\hbox{$ \stackrel{(n-1)}{\Omega_{1}}$}};
{\raise0pt\hbox{$ \stackrel{(0)}{\Omega_{2}}$}},
\ldots,{\raise0pt\hbox{$ \stackrel{(n-1)}{\Omega_{2}}$}}]
\end{equation}
which are equivalent to the three equations
\begin{equation}
[\Omega_{1},\Omega_{1}]=0=[\Omega_{2},\Omega_{2}] \qquad {\rm and}
\qquad
[\Omega_{1},\Omega_{2}]=0.
\end{equation}
As mentioned above, these last equations are equivalent to the BRST-
anti-BRST
defining equations for the derivations $ s_{1} = [\cdot,
\Omega_{1}] $ and $ s_{2} = [\cdot,\Omega_{2}] $. Thus, we have proved
the existence of the BRST-anti-BRST
transformation for any complete description of the constraint surface $
\Sigma $.
This was done not by trying to solve directly (\ref{nice1}-
\ref{nice3}), but rather
by solving the sum (\ref{familyeq}) and controlling that it splits
appropriately.

\subsection{Uniqueness of the BRST and anti-BRST generators}
\label{Subsec55}
By the standard BRST theory, the total BRST generator is unique up to
canonical transformation in the extended phase space. In its
infinitesimal
form, this result states that if $ \Omega $
and $ \Omega' $ are two nilpotent generators satisfying the same
boundary
conditions, then $ \Omega' = \Omega + [M,\Omega] $ where the function $
M $
is of ghost number zero \cite{Fish:Henneaux:Stasheff:Teitelboim}. More
explicitly,
if one decomposes $ M $ according to the resolution degree, one has
$ {\raise0pt \hbox{$ \stackrel{(r)}{\Omega'}$}} =
{\raise0pt \hbox{$ \stackrel{(r)}{\Omega}$}} + \delta
{\raise0pt \hbox{$ \stackrel{(r+1)}{M}$}} $.
Actually, one can assume that the function $ M $ is of homogeneous
ghost bidegree
$ (0,0) $, $ bigh(M)=(0,0)$.
Indeed, suppose that one has $ \Omega_{1} + \Omega_{2} =
\Omega'_{1} + \Omega'_{2} $, with the same boundary conditions. Suppose
that
until resolution degree $ p $,
\begin{eqnarray}
{\raise0pt \hbox{$ \stackrel{(r)}{\Omega'_{1}}$}} & = &
{\raise0pt \hbox{$ \stackrel{(r)}{\Omega_{1}}$}} \\
{\raise0pt \hbox{$ \stackrel{(r)}{\Omega'_{2}}$}} & = &
{\raise0pt \hbox{$ \stackrel{(r)}{\Omega_{2}}$}}, \qquad r \leq p.
\end{eqnarray}
Let us prove that there exist a canonical transformation
\begin{equation}
\Omega \rightarrow \Omega + [ {\raise0pt \hbox{$
\stackrel{(p+2)}{M}$}},
\Omega ] \label{cantransf1}
\end{equation}
 such that $ {\raise0pt \hbox{$ \stackrel{(p+1)}{\Omega'_{1}}$}} =
{\raise0pt \hbox{$ \stackrel{(p+1)}{\Omega_{1}}$}} $ and
$ {\raise0pt \hbox{$ \stackrel{(p+1)}{\Omega'_{2}}$}} =
{\raise0pt \hbox{$ \stackrel{(p+1)}{\Omega_{2}}$}} $. By construction,
one has
\begin{equation}
\delta {\raise0pt \hbox{$ \stackrel{(p+1)}{\Omega}$}} =
{\raise0pt \hbox{$ \stackrel{(p)}{D}$}} = \delta
{\raise0pt \hbox{$ \stackrel{(p+1)}{\Omega'}$}}.
\end{equation}
Thus, there exist $ {\raise0pt \hbox{$ \stackrel{(p+2)}{M}$}} $ such
that
\begin{equation}
{\raise0pt \hbox{$ \stackrel{(p+1)}{\Omega'}$}} =
{\raise0pt \hbox{$ \stackrel{(p+1)}{\Omega}$}} + \delta
{\raise0pt \hbox{$ \stackrel{(p+2)}{M}$}} \label{Transf1}.
\end{equation}
Furthermore, because $ ({\raise0pt \hbox{$ \stackrel{(p+1)}{\Omega'}$}}
-
{\raise0pt \hbox{$ \stackrel{(p+1)}{\Omega}$}}) \in {\cal K}^{*,*}_{++}
$,
one can take $ {\raise0pt \hbox{$ \stackrel{(p+2)}{M}$}} $ in
$ {\cal K}^{*,*}_{+} $, that is, $ bigh({\raise0pt \hbox{$
\stackrel{(p+2)}{M}$}})
= (0,0) $. The canonical transformation (\ref{cantransf1}) with that
solution
$ {\raise0pt \hbox{$ \stackrel{(p+2)}{M}$}} $ of (\ref{Transf1}) is the
searched-for canonical transformation. The equation (\ref{cantransf1})
splits as
\begin{eqnarray}
\Omega_{1} & \rightarrow & \Omega_{1} + [ {\raise0pt \hbox{$
\stackrel{(p+2)}{M}$}},
\Omega_{1} ] = \Omega_{1} + s_{1} {\raise0pt \hbox{$
\stackrel{(p+2)}{M}$}}
\label{spcantransf1} \\
\Omega_{2} & \rightarrow & \Omega_{2} + [ {\raise0pt \hbox{$
\stackrel{(p+2)}{M}$}},
\Omega_{2} ] = \Omega_{2} + s_{2} {\raise0pt \hbox{$
\stackrel{(p+2)}{M}$}}
\label{spcantransf2}.
\end{eqnarray}

\section{Classical BRST cohomology}
\label{Sec6}
In order to construct a gauge fixed (hamiltonian) action, it is
necessary to
define the total BRST extension $ H $ of the canonical (gauge
invariant)
hamiltonian $ H_{0} $. That is, one must find a
function $ H $ with $ gh(H) = 0 $ such that $ H = H_{0} + \cdots $ and
$ [H,\Omega] = 0 $. If one decomposes $ H $ according to the resolution
degree
\begin{equation}
H = \sum_{r = 0} {\raise0pt \hbox{$ \stackrel{(r)}{H}$}}, \qquad
res({\raise0pt \hbox{$ \stackrel{(r)}{H}$}}) = r,
\end{equation}
then, the equation $ [H,\Omega] = 0 $ is equivalent to the family of
equations
\begin{equation}
\delta {\raise0pt \hbox{$ \stackrel{(p+1)}{H}$}} =
{\raise0pt \hbox{$ \stackrel{(p)}{M}$}}[{\raise0pt \hbox{$
\stackrel{(0)}{H}$}},
\ldots, {\raise0pt \hbox{$ \stackrel{(p)}{H}$}}]
\end{equation}
where the function $ {\raise0pt \hbox{$ \stackrel{(p)}{M}$}} $ is
defined by
\cite{Henneaux:Teitelboim}
\begin{eqnarray}
{\raise0pt \hbox{$ \stackrel{(p)}{M}$}} & = & \sum_{k=0}^{p}
[ {\raise0pt \hbox{$ \stackrel{(p-k)}{H}$}},
{\raise0pt \hbox{$ \stackrel{(k)}{\Omega}$}} ]_{orig} \nonumber \\
& & \sum_{k=0}^{p} \sum_{s=0}^{k+p-1} \left\{
[ {\raise0pt \hbox{$ \stackrel{(k)}{H}$}},
{\raise0pt \hbox{$ \stackrel{(p+s+1-k)}{\Omega}$}} ]_{({\cal
P}_{A_{s}},
\eta^{A_{s}})} + [ {\raise0pt \hbox{$ \stackrel{(k)}{H}$}},
{\raise0pt \hbox{$ \stackrel{(p+s+2-k)}{\Omega}$}} ]_{(\lambda_{A_{s}},
\pi^{A_{s}})} \right\} \label{MPexp}
\end{eqnarray}
The general theorems of BRST theory guarantee the existence of $ H $.
Again, one
has here a stronger result, namely, $ H $ can be chosen to be of ghost
bidegree
$ (0,0) $.
Clearly, one has $ {\raise0pt \hbox{$ \stackrel{(0)}{H}$}} = H_{0} $.
It is also
easy to see that
{\footnote As usual, we define $ [H_{0},G_{A_{0}}] = V^{B_{0}}_{A_{0}}
G_{B_{0}}$
; that is the first class condition on the canonical hamiltonian $
H_{0} $.}
$ {\raise0pt \hbox{$ \stackrel{(1)}{H}$}} =
\ghost{.5}{-1}{0}{\cal P}{A}{0} V^{A_{0}}_{B_{0}}
\cghost{0}{1}{0}{\eta}{A}{0}
+ \ghost{.5}{0}{-1}{\cal P}{A}{0} V^{A_{0}}_{B_{0}}
\cghost{0}{0}{1}{\eta}{A}{0} $.
This shows that $ bigh({\raise0pt \hbox{$ \stackrel{(0)}{H}$}}) =
bigh({\raise0pt \hbox{$ \stackrel{(1)}{H}$}}) = (0,0) $. As in lemma
\ref{PositiveD}, one can conclude that $ {\raise0pt \hbox{$
\stackrel{(1)}{M}$}} $
in (\ref{MPexp}) belongs to $ {\cal K}^{*,*}_{+} $. Because
$ \delta {\raise0pt \hbox{$ \stackrel{(1)}{M}$}} = 0 $, by
theorem \ref{Ther52}, there exists $ {\raise0pt \hbox{$
\stackrel{(2)}{H}$}} $
such that $ bigh({\raise0pt \hbox{$ \stackrel{(2)}{H}$}}) = (0,0) $ and
$ \delta {\raise0pt \hbox{$ \stackrel{(2)}{H}$}} =
{\raise0pt \hbox{$ \stackrel{(1)}{M}$}} $. Continuing
in the same fashion, one finally obtains the following theorem
\newtheorem{BRSTham}{Theorem}[section]
\begin{BRSTham}
\label{BRSTham}
The total BRST invariant extension $ H $ of the canonical hamiltonian $
H_{0} $
may be chosen in such a way that $ bigh(H) = (0,0) $. The equation $
[H,\Omega] = 0 $
imply then that $ H $ is both BRST and anti-BRST invariant,
that is, $ s_{1}H = 0 = s_{2}H $.
\end{BRSTham}
By standard BRST arguments one also obtains easily the
\newtheorem{Uniqueham}[BRSTham]{Theorem}
\begin{Uniqueham}
\label{Uniqueham}
The total BRST extension $ H $ of the canonical hamiltonian $ H_{0} $
is
unique up to BRST-exact term : the equations $ [H,\Omega] = 0 =
[H',\Omega] $,
with $ {\raise0pt \hbox{$ \stackrel{(0)}{H}$}} =
{\raise0pt \hbox{$ \stackrel{(0)}{H'}$}} = H_{0} $ imply the existence
of
a function $ K $ such that $ H = H' + [K,\Omega] $.
\end{Uniqueham}
The {\em gauge fixed hamiltonian} $ H_{\Psi} = H + s\Psi $ is simply a
choice
of a representant in the equivalence class of BRST invariant extensions
of the canonical hamiltonian $ H_{0} $.

\section{Comparison with the standard BRST formalism}
\label{Sec7}
It is clear that the above approach yields the same physical results as
the standard BRST formalism. Indeed, it is known that these physical
results
do not depend on the particular resolution of $ C^{\infty}(\Sigma) $
that is
adopted. However, it is of interest to make a more explicit contact
with the
standard BRST construction. To that end, we observe that the
BRST generator $ \Omega_{1} $ given here starts as
\begin{equation}
\label{defBRST}
\Omega_{1} = - \sum_{k=0} \cghost{0}{k+1}{0}{\eta}{A}{k} \delta_{1}
\ghost{.5}{-(k+1)}{0}{\cal P}{A}{k} + ``{\rm more}", \qquad
[\Omega_{1},
\Omega_{1}] = 0
\end{equation}
where the operator $ \delta_{1} $ provides a homological resolution of
the
algebra $ C^{\infty}(\Sigma) $. The equations (\ref{defBRST}) precisely
define
the standard BRST of the standard theory charge with a non minimal
sector :
besides the minimal variables $ \ghost{.5}{-(k+1)}{0}{\cal P}{A}{k} $
and
$ \cghost{0}{k+1}{0}{\eta}{A}{k} $, there are extra non minimal
variables (all
the others). Hence, we can indeed identify $ \Omega_{1} $ with the
standard
(non minimal) BRST generator. The ghosts $ \cghost{0}{1}{1}{\pi}{A}{0}
$, which
appear in our approach as ghosts of ghosts related to the duplication
of the constraints,
are viewed as non minimal variables in the standard BRST context. Note
that this
non minimal sector turns out to be the non minimal sector required for
convenient gauge fixing (for instance, the Feynman gauge for the Yang-
Mills
action).

The ghost number of the standard BRST formalism can be
expressed as
\begin{equation}
gh_{standard} = gh_{1} - gh_{2}.
\end{equation}
Hence, one has $ gh_{standard}(\Omega_{1}) = +1 $ and
$ gh_{standard}(\Omega_{2}) = -1 $. Moreover, the total BRST extension
of the
canonical hamiltonian is also a standard BRST extension for the
standard BRST
charge $ \Omega_{1} $ : $ [H,\Omega_{1}]=0 $.
The ambiguity in $ H $ explained in theorem \ref{Uniqueham} may be
rewritten
as $ H \rightarrow H + [K',\Omega_{1}] $ where $ K' $ is such that
$ [K',\Omega_{1}] $ is anti-BRST invariant. Thus, it is of the standard
form from the BRST point of view based on
$ \Omega_{1} $. Indeed, because $ bigh(sK) = (0,0) $, one has that $ sK
$ is
BRST and anti-BRST invariant. On the other hand, $ sK|_{{\cal P} =
\lambda = G
= 0} = 0 $. Thus, $ sK $ is ({\em i}) $s_{1} $-closed and ({\em ii}) an
extension of zero. Hence, it is $ s_{1} $-exact (see
\cite{Henneaux:Teitelboim}),
$ sK = s_{1}K' $ for some $ K' $ with $ s_{1}K' $ anti-BRST invariant.

Actually, from the standard BRST viewpoint, one only requires the
standard
ghost number of the BRST extension of $ H_{0} $ to be zero, i.e., $ H $
may contain also terms of bidegree $ (k,k) $ with $ k \not= 0 $.
We have the following general theorem that allows one to make the link
between the
standard BRST theory and the BRST-anti-BRST theory at the gauge fixing
level:
\newtheorem{Gaugefixing}{Theorem}[section]
\begin{Gaugefixing}
\label{Gaugefixing}
Let $ \Psi $ be a fermionic function such that $ s\Psi $ contains
only terms of ghost bidegree of the form $ (k,k) $. Then
$ s\Psi $ is BRST and anti-BRST invariant and $ s\Psi = s_{1}\Psi'$ for
some fermionic function $ \Psi' $. Conversely, if
$ s_{1}\Psi' $ is anti-BRST invariant and contains only terms of ghost
bidegree of the form $ (k,k) $, then it can be written as $ s\Psi $ for
some fermionic function $ \Psi $.
\end{Gaugefixing}
{\bf Proof of theorem \ref{Gaugefixing}}:
Let us expand the function $ \Psi $ according to the standard ghost
number~: $ \Psi
= \sum_{n} \Psi_{n} $, where $ gh_{standard}(\Psi_{n}) = n $. The
requirement that
$ gh_{standard}((s_{1} + s_{2}) \Psi) = 0 $ translates into the
following familly of
equations :
\begin{eqnarray}
(s_{1} + s_{2}) \Psi & = & s_{1} \Psi_{-1} + s_{2} \Psi_{1} , \\
s_{2} \Psi_{-1} + s_{1} \Psi_{-3} & = & 0, \\
s_{1} \Psi_{1} + s_{2} \Psi_{3} & = & 0 , \\
\vdots \nonumber
\end{eqnarray}
Hence, we have $ s_{1}(s_{1} \Psi_{-1} + s_{2} \Psi_{1}) = s_{1}s_{2}
\Psi =
- s_{2}s_{1} \Psi_{1} = s^{2}_{2} \Psi_{3} = 0 $, and similarly, one
can see
that $ s_{2} (s_{1} \Psi_{-1} + s_{2} \Psi_{1}) = s^{2}_{1} \Psi_{-3} =
0 $.
One can also see that $ s_{2} \Psi_{1} $ is $s_{1}-$exact, because it
is
$ s_{1}-$closed, of standard ghost number zero and it vanishes when $
{\cal P} = \lambda
= G = 0 $. Thus, one finds that
\begin{equation}
(s_{1} + s_{2}) \Psi = s \Psi' ,
\end{equation}
where the function $ \Psi' $ is of standard ghost number minus one and
such that
$ s_{2}s_{1} \Psi' = 0 $.

Conversely, suppose that one has $ s_{1} \Psi' $ with $ s_{2}s_{1}
\Psi' = 0 $ and
$ gh_{standard}( \Psi') = -1 $. Then, one can find $ \Psi $ such that $
s_{1} \Psi' =
(s_{1} + s_{2}) \Psi $. Indeed, one has the relation $ s_{1} \Psi' =
(s_{1} + s_{2}) \Psi' - s_{2} \Psi' $. But $ s_{1}s_{2} \Psi' = 0 $ and
$ s_{2} \Psi' $ is of standard ghost number -2 . Hence, $ s_{2} \Psi' $
is
$ s_{1}-$trivial (no $ s_{1}$-cohomology at standard negative ghost
degrees)~:
$ s_{2} \Psi' = - s_{1} \Psi_{-3} $ and so one obtains
$ \Psi' = (s_{1} + s_{2}) (\Psi' + \Psi_{-3}) - s_{2} \Psi_{-3} $.
Going on
recursively in the same fashion at lower standard ghost numbers, one
conclude that
$ s_{1} \Psi' = ( s_{1} + s_{2} ) \Psi $.  {\bf QED}

Those gauge fixed hamiltonians are to be used in the path integral in
order
to quantize gauge systems. The fact that the path integral does not
depend on the choice of the fermionic function $ \Psi $ follows from
the
Fradkin and Vilkovisky theorem \cite{FradVil}. On the other hand, the
path integral obtained by applying the BRST-anti-BRST formalism is of
the form
of the standard BRST path integral, since $ s_{1} \Psi' = s \Psi$.
Hence, the equivalence of the BRST-anti-BRST
formalism with the standard BRST formalism (at the path integral level)
is
obvious.

\section{Comparison with the $ sp(2) $ formalism}
\label{Sec8}
The $ sp(2) $ formalism has attracted considerable attention in
connection with
string field theory, see
\cite{Hoyosetal,BatLav1,BatLav2,BatLav3,BatLav4,Aratyn,BauSieZwi,SieZ
wi
}. Our
BRST-anti-BRST formulation reproduces the $ sp(2) $ formulation of
\cite{BatLav3,BatLav4} when the ambiguity in $ \Omega $ is
appropriately handled.
This can be seen as follows. First of all, the spectra of ghosts and of
ghost momenta are the same. Using the notations of \cite{BatLav2}, one
has
the following correspondence for the ghost momenta :
\begin{eqnarray}
& & \left\{
\begin{array}{lll}
\ghost{.5}{-1}{0}{\cal P}{A}{0} & \longleftrightarrow & {\cal
P}_{A_{0}|1} \\
\ghost{.5}{0}{-1}{\cal P}{A}{0} & \longleftrightarrow & {\cal
P}_{A_{0}|2}
\end{array} \right. \\
& & \ghost{0}{-1}{-1}{\lambda}{A}{0} \longleftrightarrow
\lambda_{A_{0}} \\
& & \left\{
\begin{array}{lll}
\ghost{.5}{-2}{0}{\cal P}{A}{1} & \longleftrightarrow & {\cal
P}_{A_{1}|11} \\
\ghost{.5}{-1}{-1}{\cal P}{A}{1} & \longleftrightarrow & {\cal
P}_{A_{1}|12}
\equiv {\cal P}_{A_{0}|21} \\
\ghost{.5}{0}{-2}{\cal P}{A}{1} & \longleftrightarrow & {\cal
P}_{A_{1}|22}
\end{array} \right. \\
& & \vdots \nonumber \\
& & \ghost{.5}{-i}{-j}{\cal P}{A}{k} \longleftrightarrow
{\cal P}_{A_{k}|\underbrace{\scriptstyle{1 \ldots 1}}_{i}
\underbrace{\scriptstyle{2 \ldots 2}}_{j}} \\
& & \vdots \nonumber \\
& & \ghost{0}{-i}{-j}{\lambda}{A}{k} \longleftrightarrow
\lambda_{A_{k}|\underbrace{\scriptstyle{1 \ldots 1}}_{i}
\underbrace{\scriptstyle{2 \ldots 2}}_{j}} \\
& & \vdots \nonumber
\end{eqnarray}
where $ {\cal P}_{A_{k}|a_{1} \ldots a_{k+1}} $ and $
\lambda_{A_{k}|a_{1}
\ldots a_{k}} $ are symmetric $ sp(2) $ tensors. The identification for
the
ghosts are then obvious. Second, the ghost number $ gh $
introduced in the present paper is exactly the {\em new ghost number}
of
\cite{BatLav3,BatLav4}. Finally, a close inspection of the equations
(\ref{nice1}),
(\ref{nice2}) and (\ref{nice3}) shows that one can make the choice
$ S \Omega_{1} = \Omega_{2} $ and $ S \Omega_{2} = \Omega_{1} $ (this
follows
from the fact that $ S D_{11} = D_{22} $, $ S D_{22} = D_{11} $
$ S D_{12} = D_{21} $ and theorem \ref{Ther42}). Then one has
\begin{eqnarray}
S s_{1} S & = & s_{2} \\
S s_{2} S & = & s_{1}
\end{eqnarray}
and
\begin{equation}
S s S = s
\end{equation}
With that choice, there is a complete symmetry between the BRST and the
anti-BRST
generators, as in the $ sp(2) $ theory and the generators $ \Omega_{1}
$ and $ \Omega_{2} $
coincide with the generators $ \Omega^{a} $ ( $ a = 1,2 $ ) of
references
\cite{BatLav3,BatLav4}.

\section{Conclusions}
\label{Sec9}
In this paper we have explored the algebraic structure of the BRST-
anti-BRST
formalism. We have proven the existence of the BRST-anti-BRST
transformation
for an arbitrary gauge system. To that end, it was found necessary to
enlarge the
ghost system and to introduce a Koszul-Tate {\em biresolution} of the
algebra of
smooth functions defined on the constraint surface. One can then apply
the
{\em standard BRST techniques} to the corresponding reducible
description of the constraint
surface, to get directly the generator $ \Omega $ of the sum of the
BRST and the
anti-BRST transformations. A crucial positivity theorem controls that
$ \Omega $ indeed splits as a sum of just two terms ($\Omega^{{\rm
BRST}} = \Omega_{1}$
and $ \Omega^{{\rm anti-BRST}} = \Omega_{2} $), and no more. This
positivity
theorem, in turn, is a consequence of the algebraic properties of the
Koszul-Tate
biresolution. Our approach clearly explains the complexity of the
ghost-antighost
spectrum necessary for the BRST-anti-BRST formulation and also shows in
a
straightformard way the equivalence between the
standard BRST formalism and the BRST-anti-BRST one. The arguments
developped
in this article can be applied, with some modifications, to the
extended antifield-antibracket
formalism. We shall return to this question in a separate publication
\cite{Gregoire:Henneaux2}.
\vfill
\pagebreak

\end{document}